\documentclass[10pt,journal,compsoc]{IEEEtran}

\usepackage{cite}             
\usepackage{hyperref}
\usepackage{amsmath,amssymb}
\usepackage{graphicx}
\usepackage{array}
\usepackage[caption=false,font=footnotesize]{subfig}
\usepackage{booktabs}
\usepackage{multirow}
\usepackage[table]{xcolor}
\usepackage{mathrsfs}
\usepackage[linesnumbered,ruled,vlined]{algorithm2e}

\title{All Cities are Equal: A Unified Human Mobility Generation Model Enabled by LLMs}

\author{
    Bo~Liu,
    Tong~Li,~\IEEEmembership{Member,~IEEE},
    Zhu~Xiao,~\IEEEmembership{Senior Member,~IEEE},\\
    Ruihui~Li,
    Geyong~Min,~\IEEEmembership{Member,~IEEE},
    Zhuo~Tang,~\IEEEmembership{Member,~IEEE},
    Kenli~Li,~\IEEEmembership{Senior Member,~IEEE}

    \IEEEcompsocitemizethanks{\IEEEcompsocthanksitem B.~Liu, T.~Li, Z.~Xiao, R.~Li, Z.~Tang, and K.~Li are with the College of Computer Science and Electronic Engineering, Hunan University, Changsha 410082, China. 
    \IEEEcompsocthanksitem G.~Min is with the Department of Computer Science, University of Exeter, UK.}  
}
 
\begin{document}
\IEEEtitleabstractindextext{
\begin{abstract}
Synthetic human mobility generation is gaining traction as an ethical and practical approach to supporting the data needs of intelligent urban systems. Existing methods perform well primarily in data-rich cities, while their effectiveness declines significantly in cities with limited data resources. However, the ability to generate reliable human mobility data should not depend on a city's size or available resources—all cities deserve equal consideration.
To address this open issue, we propose UniMob, a unified human mobility generation model across cities. UniMob is composed of three main components: an LLM-powered travel planner that derives high-level, temporally-aware, and semantically meaningful travel plans; a unified spatial embedding module that projects the spatial regions of various cities into a shared representation space; and a diffusion-based mobility generator that captures the joint spatiotemporal characteristics of human movement, guided by the derived travel plans. We evaluate UniMob extensively using two real-world datasets covering five cities.
Comprehensive experiments show that UniMob significantly outperforms state-of-the-art baselines, achieving improvements of over 30\% across multiple evaluation metrics. Further analysis demonstrates UniMob's robustness in both zero- and few-shot scenarios, underlines the importance of LLM guidance, verifies its privacy-preserving nature, and showcases its applicability for downstream tasks. The code is available at: \url{https://anonymous.4open.science/r/UniMob-E544}.
\end{abstract}
\begin{IEEEkeywords}
Human mobility generation, diffusion models, large language models, cross-city generalization, privacy
\end{IEEEkeywords}
}

\maketitle

\IEEEdisplaynontitleabstractindextext

\section{Introduction}

With the rapid growth of urbanization and an increasing global emphasis on sustainable development, cities worldwide are striving to build more intelligent and efficient urban systems~\cite{Luca2021SurveyDLMobility, zheng2014urban}. Key objectives in this pursuit include enhancing transportation efficiency~\cite{wei2018intellilight, liu2020polestar}, optimizing land use~\cite{zhang2018simultaneous, zheng2023road}, and improving the coordination of urban infrastructure resources~\cite{zhang2023deep, li2023artificial}. Achieving these goals relies heavily on the availability and analysis of human mobility data~\cite{xu2023urban, nilforoshan2023human, bassolas2019hierarchical}, which provides an essential foundation for both research and practical applications in urban environments. However, collecting large-scale, high-quality individual mobility data presents significant challenges~\cite{liu2025vehicle, pellungrini2017data}. The process is often expensive and complicated by strict privacy concerns and restrictions that limit data collection and sharing. In response to these obstacles, the generation of synthetic human mobility data has emerged as a promising alternative~\cite{GANMobility2020, Liu2024Act2Loc, Yuan2024GenDailyActivities}. This approach enables the creation of synthetic human mobility datasets that accurately reflect the essential characteristics of real-world movement patterns while effectively safeguarding individual privacy. As a result, synthetic human mobility generation is rapidly emerging as a viable and ethical solution for meeting the functional requirements of data-driven intelligent urban systems.

Existing studies have devoted considerable effort to human mobility generation. Early sequence-based methods, such as Markov models and RNN variants~\cite{feng2018deepmove}, focused on capturing short-term temporal dependencies in movement histories. GAN-based approaches, including MoveSim~\cite{feng2020learning} and SeqGAN~\cite{yu2017seqgan}, began modeling higher-order spatiotemporal patterns to generate human mobility. More recently, diffusion-based models such as TrajGDM~\cite{chu2024simulating} and CoDiffMob~\cite{zhang2025noise} have framed mobility synthesis as conditional denoising diffusion over latent representations. 
However, existing methods face significant constraints in several respects. First, these models depend heavily on large volumes of historical trajectory data for training~\cite{yuan2022activity, Wang2024cola}, which limits their applicability in data-scarce or privacy-sensitive settings, such as cities with limited sensing infrastructure or emerging urban areas with minimal sensors or travel surveys~\cite{feng2020learning, wang2019urban}. Moreover, because these methods typically encode information such as spatial layouts, Points of Interest (POI) distributions, and local movement frequencies, the trained models are often tightly coupled to the spatial configuration and mobility dynamics of the data source city~\cite{Wang2024cola}. As a result, the performance of existing methods tends to be strong in data-rich cities but degrades noticeably in low-resource contexts~\cite{zhang2025noise}. Notably, \emph{all cities are equally important-reliable human mobility generation is crucial regardless of city size or resources}. Applications such as transportation planning, emergency simulation, and urban resource allocation demand reliable synthetic mobility data in both large, resource-rich cities and data-scarce environments.

\begin{figure*}[tb]
    \centering
    \vspace{-3mm}
    \subfloat[Beijing]{\includegraphics[width=0.45\linewidth]{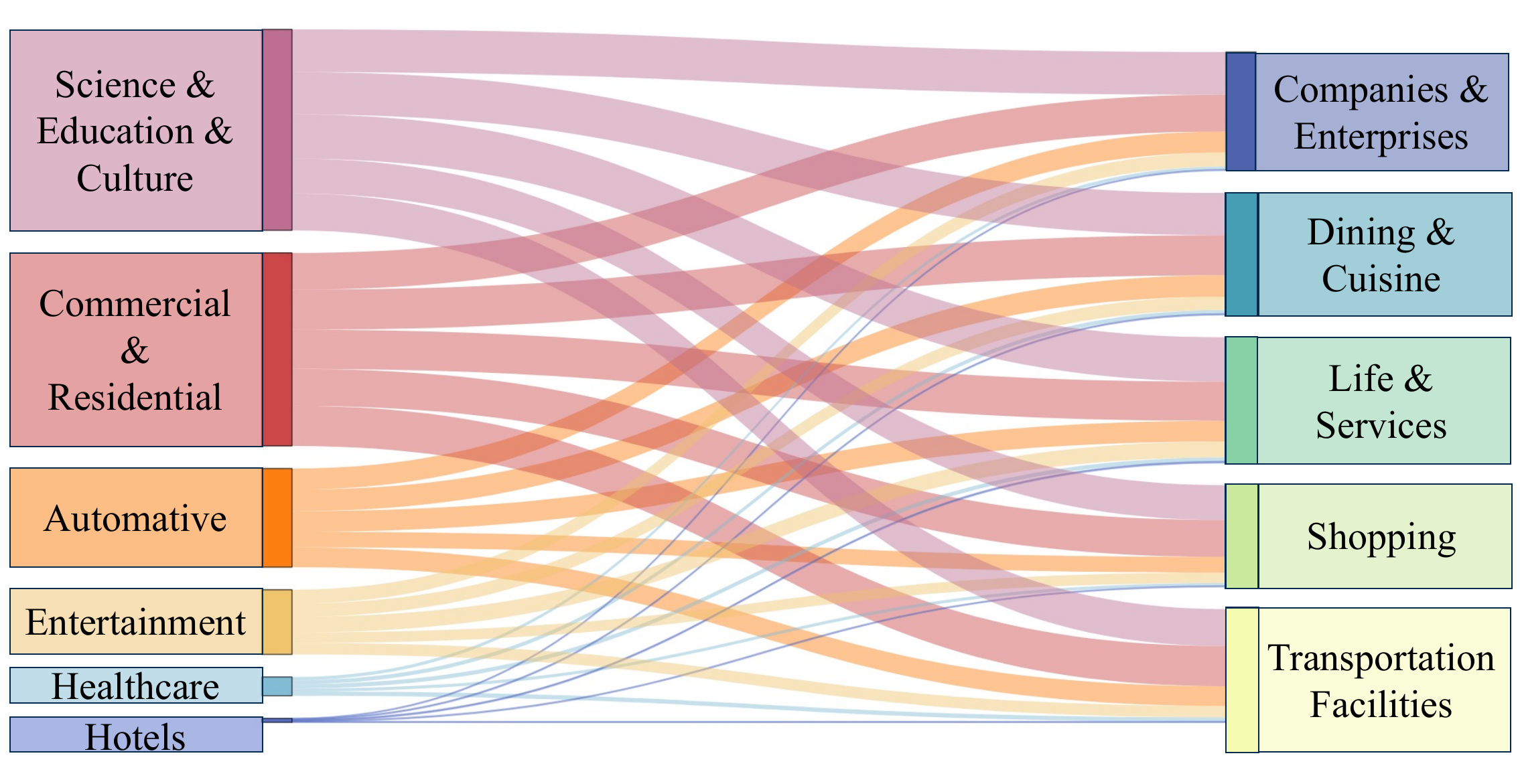}}
    \subfloat[Shanghai]{\includegraphics[width=0.45\linewidth]{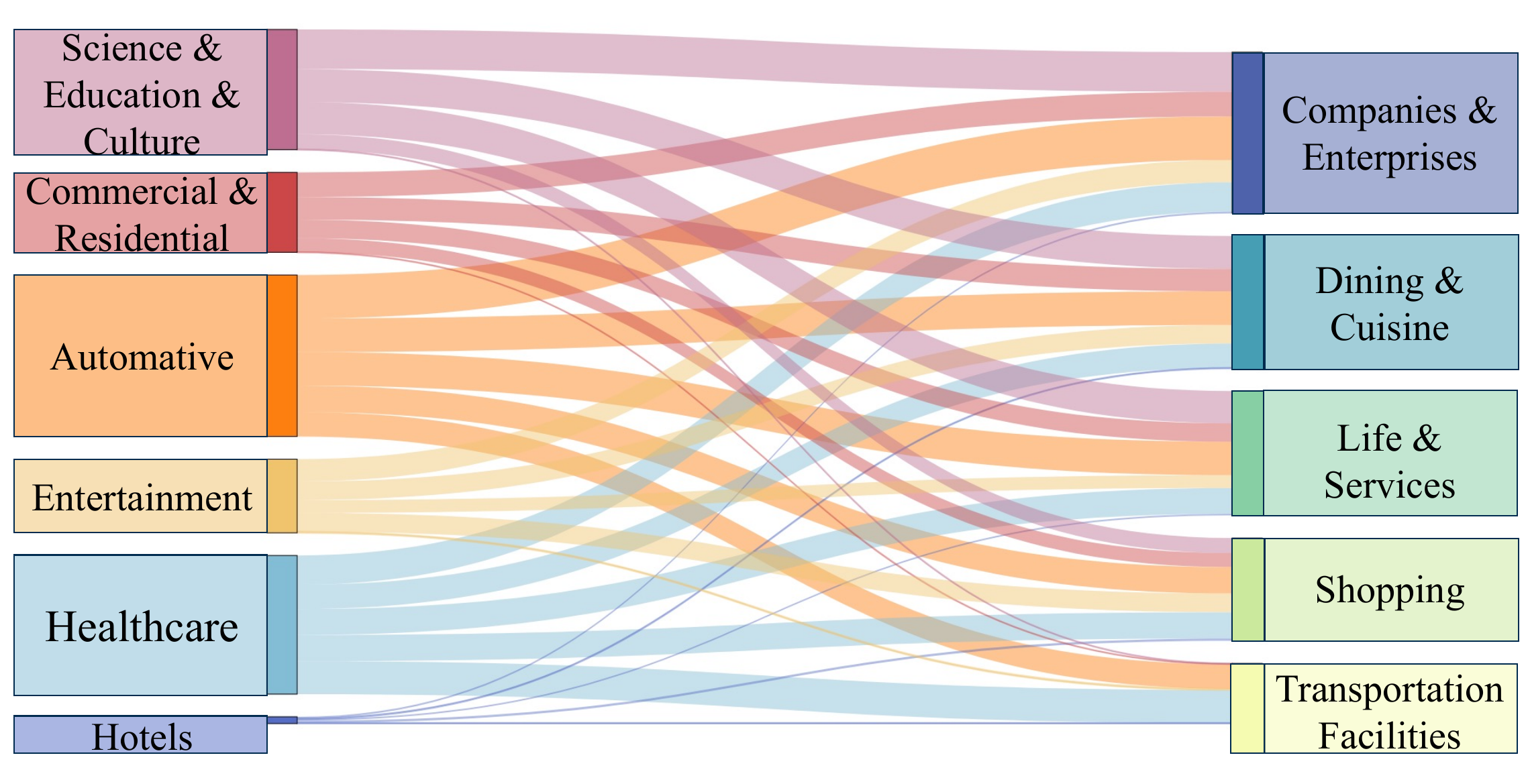}}
    \vspace{-2mm}
    \caption{Visualization of POI transitions analysis based on mobile phone datasets from Beijing and Shanghai. Similar transition patterns from departure (left) to destination(right) in Beijing and Shanghai can be observed.}
    \vspace{-2mm}
    \label{fig:dataset_visualization}
\end{figure*}

In this paper, we aim to propose a unified generative framework designed to generate synthetic human mobility data, enabling more equitable support for downstream mobility analysis and decision-making across cities.
Achieving this objective, however, presents several significant challenges.
\textbf{Firstly}, cities differ greatly in their urban environments, exhibiting unique spatial layouts and POI distributions, leading to highly diverse mobility dynamics from one city to another~\cite{yuan2012discovering}. Such inter-city variability makes it difficult to generate human mobility across cities. Thus, it is crucial, while challenging to capture the fundamental mechanisms that drive human mobility, abstracting the high-level intentions behind when, where, and why people move, regardless of the particularities of any single city.
\textbf{Secondly}, cities often employ different spatial partitioning schemes and exhibit diverse spatial structures~\cite{Wang2025GTG}. A truly unified mobility generation framework must leverage data from all cities by relying on shared, unified representations. Bridging the unique spatial configurations of individual cities into a unified representation space is both crucial and challenging. \textbf{Thirdly}, although human mobility often follows consistent patterns, such as commuting to work in the morning and returning home in the evening, the actual movement behaviors are strongly influenced by local environments, regional functions, and external factors. This diversity becomes especially pronounced when modeling mobility across different cities, adding to the variability of human movement. Therefore, generating human mobility across cities requires capturing not only the shared regularities but also the probabilistic nature of behavioral diversity.

To address the challenges outlined above, we propose UniMob, a unified generative framework for high-fidelity synthetic human mobility generation across cities. UniMob comprises three key components: (i) an LLM-based travel planner for inferring high-level semantic travel intent, (ii) a unified spatial embedding that maps diverse spatial structures into a common representation space, and (iii) a diffusion-based mobility generator for modeling realistic and stochastic human movement. 
Specifically, LLM-based travel planner addresses the \textbf{first} challenge, by inferring both the temporal mode and semantic intent of travel to guide the generation process. By reasoning over structured contextual inputs, such as current time and POI distribution of location, the LLMs are fine-tuned to derive a high-level temporal and semantic travel plan driven by travel intent. These semantic travel plans reveal that fundamental mobility mechanisms are independent of specific spatial layouts or geographic features across cities.
Next, the unified spatial embedding module tackles the \textbf{second} challenge. A spatiotemporal encoder embeds the varying spatial structures of different cities into a unified spatial representation space. For generalization, a lightweight tunable decoder maps the unified spatial representations back onto the specific locations of individual cities. Finally, a diffusion-based mobility generator addresses the \textbf{third} challenge. As a probabilistic generative framework, the diffusion process models the joint underlying spatiotemporal distribution of human mobility across cities, guided by high-level temporal and semantic travel plans.

The main contributions can be summarized as follows:

$\bullet$ We emphasize the importance of developing a generative framework that treats all cities with equal significance. Our goal is to investigate a unified generative framework capable of producing synthetic human mobility data, enabling more equitable support for downstream mobility analysis and decision-making across cities.

$\bullet$ We propose UniMob, a unified human mobility generation model. UniMob consists of three key components: an LLM-based travel planner that derives high-level temporal and semantic travel plans driven by user intent; a unified spatial embedding module, which maps the diverse spatial structures of different cities into a shared spatial representation space; a diffusion-based mobility generator that models the joint spatiotemporal distribution of human mobility across cities, guided by the high-level travel plans.

$\bullet$ We evaluate UniMob extensively using two real-world datasets covering five cities. Our results demonstrate that UniMob outperforms state-of-the-art baselines by more than 30\% across multiple metrics. Further analysis highlights its effectiveness in zero-shot and few-shot scenarios, emphasizes the value of LLM guidance, confirms its privacy-preserving capabilities, and demonstrates its utility for various downstream tasks.

The rest of this paper is organized as follows. Section~\ref{preliminaries} presents preliminaries, including human mobility modeling, an overview of Large Language Models, and the problem definition. Section~\ref{methods} details the UniMob framework, comprising its travel planner, unified spatial embedding, and mobility generator. Section~\ref{experiments} describes experimental setups and presents comprehensive results. Section~\ref{related} discusses related work, and Section~\ref{conclusion} concludes the paper.
\section{Preliminaries}
\label{preliminaries}

\begin{figure*}[tb]
    \centering
    \vspace{-2mm}
    \includegraphics[width=1\linewidth]{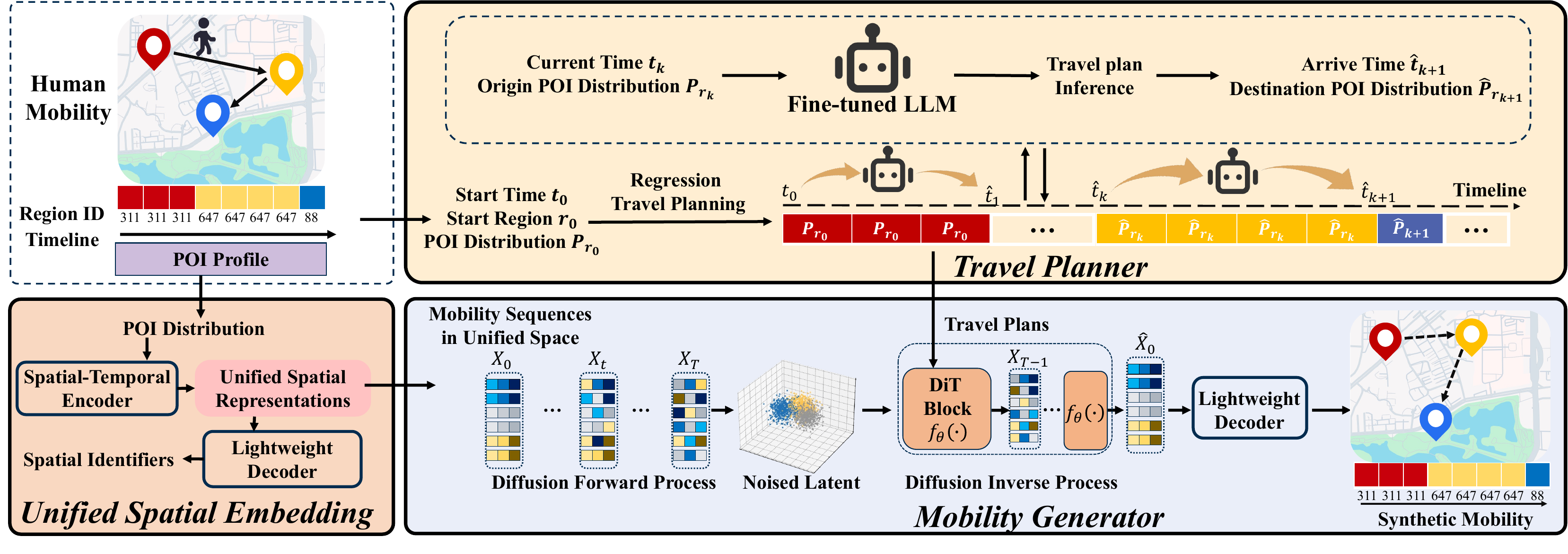}
    \vspace{-2mm}
    \caption{Framework overview of \textbf{UniMob}. Based on a start time $t_0$ and POI distribution of a start region$\mathbf{p}_{r_0}$, the travel planner uses a fine-tuned LLM to infer temporal and semantic travel plans driven by individual intent. Unified spatial embedding maps diverse spatial structures into unified space, enabling human mobility generation across cities. Mobility generator synthesizes human mobility through a denoising process conditioned on travel plans.}
    \vspace{-2mm}
    \label{fig:framework}
\end{figure*}

\subsection{Human Mobility Modeling}

Human mobility refers to the spatial and temporal movement behavior of individuals, typically characterized by sequences of visited locations over time. Understanding and modeling such mobility behaviors is essential for a wide range of applications, including urban planning, transportation management, public health monitoring, and location-based services~\cite{Solmaz2019SurveyMobility}. Mathematically, human mobility can be expressed as a spatial-temporal $S=[s_1,s_2,...,s_n]$, where the $i$-th element $s_i = (r_i, t_i)$ denotes a spatial-temporal record consisting of a visited location $r_i$ and the corresponding timestamp $t_i$. In our case, $r_i$ denotes a discrete region identifier based on Voronoi partitioning, and $t_i$ is the time of the visit.

Although human mobility varies across cities in terms of spatial layouts and routes, it often follows unified semantic patterns driven by intent and functional demand. We conduct a cross-city region-level POI transitions analysis based on mobile phone location datasets from Beijing and Shanghai. For each region, we assign a semantic label based on the dominant POI categories and analyze how user movements traverse these regions. As shown in Figure~\ref{fig:dataset_visualization}, despite differences in spatial layouts and POI distributions between Beijing and Shanghai, the two cities exhibit semantically similar mobility patterns, following common routines such as commuting, dining, and accessing services. Therefore, human mobility is intent-driven and follows consistent semantic patterns across cities.

\subsection{Large Language Model}
Large Language Models (LLMs), which are trained on massive web-scale corpora, possess extensive commonsense and domain-specific knowledge about human activities and urban functionality~\cite{wu2025survey}. This capability makes LLMs a powerful tool for unified human mobility generation across cities, particularly due to their ability to infer semantic travel intents~\cite{gong2024mobility}. By leveraging their deep understanding of both general and context-specific patterns, LLMs can accurately model and predict human movement behaviors in diverse urban environments. As a result, LLMs hold significant promise for transforming urban mobility research and supporting the development of smarter, more efficient cities.

\subsection{Problem Definition}
Rather than building separate models for each urban context, we aim to learn a unified generation framework that produces human mobility regardless of the city's geographic or structural differences. Mathmatically, given a set of real-world human mobility \(S = [s_1,s_2,...,s_n]\), where each \(S_i = [(r^i_0, t^i_0), (r^i_1, t^i_1), (r^i_2, t^i_2), \dots, (r^i_K, t^i_K)] \), the objective of the human mobility generation is to learn a generative model $\mathcal{G}$, that can generate synthetic human mobility \(\hat{S} = [\hat{s}_1,\hat{s}_2,...,\hat{s}_{n'}]\), where each \( \hat{s_i} = [(\hat{r}^i_0, \hat{t}^i_0), (\hat{r}_1, \hat{t}^i_1), (\hat{r}_2, \hat{t}^i_2), \dots, (\hat{r}_K, \hat{t}_K)] \). The generated synthetic mobility preserves the spatiotemporal characteristics of real human mobility and maintains the utility for downstream applications without revealing sensitive information about individuals associated with the real human mobility $S$. Besides, the generative model $\mathcal{G}$ can generate mobility across cities through a unified paradigm, enabling generalization to a new city without retraining the entire model.

\section{UniMob}
\label{methods}

\subsection{Framework Overview}
Figure~\ref{fig:framework} illustrates the framework overview of UniMob. Specifically, UniMob consists of three core components:
(1) \textbf{Travel planner}: this module fine-tunes a LLM to infer high-level travel plans. By reasoning over the instruction constructed by the current time and origin POI distribution, the travel planner infers both the temporal mode
and the semantic intent of travel, and these travel plans serve as conditions for the diffusion process that follows;
(2) \textbf{Unified spatial embedding}: this module leverages a spatial-temporal encoder to embed diverse spatial structure into unified representations. During inference, this module uses a lightweight decoder to map generated unified mobility representations back to city-specific spatial structures; In this case, diverse spatial structures can be processed equally in a unified manner, supporting consistent and equitable data synthesis across cities;
(3) \textbf{Mobility generator}: this module models probabilistic distribution of human mobility and generates mobility representations by progressively denoising mobility representations under the temporal mode and semantic intent guidance.

\subsection{Travel Planner}
Urban spatial structures, functional zones, and temporal activity rhythms, which vary across cities, play a crucial role in shaping individual travel intent and plans, thereby influencing overall human mobility patterns. To extract the travel semantic intent of human mobility, we propose an LLM-based \textbf{travel planner}. Leveraging the extensive commonsense and knowledge about human activities of a fine-tuned LLM, the travel planner infers high-level temporal and semantic travel plans by reasoning over urban contextual inputs. These travel plans, decoupled from specific spatial configurations, represent travel intent in a unified manner and can guide mobility generation. 

We design an autoregressive reasoning process in which the LLM models mobility as a sequence of arrival events. As illustrated in Algorithm~\ref{alg:plan_algorithm}, in each recursive step, a fine-tuned LLM is used to infer next arrival time $\hat{y}_{time}$ and destination POI distribution $\hat{y}_{poi}$ based on current time $t$ and origin POI distribution of departure region $p$. Then, the previous current time and origin POI distribution is updated with the inferred values, recurssively planning the temporal and semantic travel plans, constructed with $\mathbf{\hat{M}}$ and $\mathbf{\hat{D}}$, until it reaches the max mobility length $T_{max}$.

\begin{algorithm}[tb]
\label{alg:plan_algorithm}
\caption{Travel Planning Algorithm}
\KwIn{Start time $t_0$, start region $r_0$, POI distribution of start region $\mathbf{p}_{r_0}$; mobility length $K$, time slot interval $\Delta t$}
\KwOut{Predicted temporal travel plan $\hat{\mathbf{M}} = [\hat{t}_1, \dots]$ and semantic travel plan $\hat{\mathbf{D}} = [\hat{\mathbf{p}}_1, \dots]$}

Initialize: $\hat{\mathbf{t}} \leftarrow []$, $\hat{\mathbf{p}} \leftarrow []$ \;
Set current time $t \leftarrow t_0$, current POI vector $\mathbf{p} \leftarrow \mathbf{p}_{r_0}$ \;
Set maximum allowed time: $T_{\text{max}} \leftarrow t_0 + K \cdot \Delta t$ \;

\While{$t \leq T_{\text{max}}$}{
    Formulate prompt $p$ using $(t, \mathbf{p})$ \;
    $\mathbf{h} \leftarrow \text{LLM}(p)$ \; 
    $[\hat{y}_{\text{time}}, \hat{y}_{\text{poi}}] \leftarrow \text{OutputLayer}(\mathbf{h})$ \;
    Append $\hat{y}_{\text{time}}$ to $\hat{\mathbf{M}}$ \; 
    Append $\hat{y}_{\text{poi}}$ to $\hat{\mathbf{D}}$ \;
    Update $t \leftarrow \hat{t}_{\text{latest}}$, $\mathbf{p} \leftarrow \hat{\mathbf{p}}_{\text{latest}}$ \;
}
\Return{$\hat{\mathbf{M}}, \hat{\mathbf{D}}$}
\end{algorithm}

\begin{figure}
    \centering
    \vspace{-3mm}
    \includegraphics[width=1\linewidth]{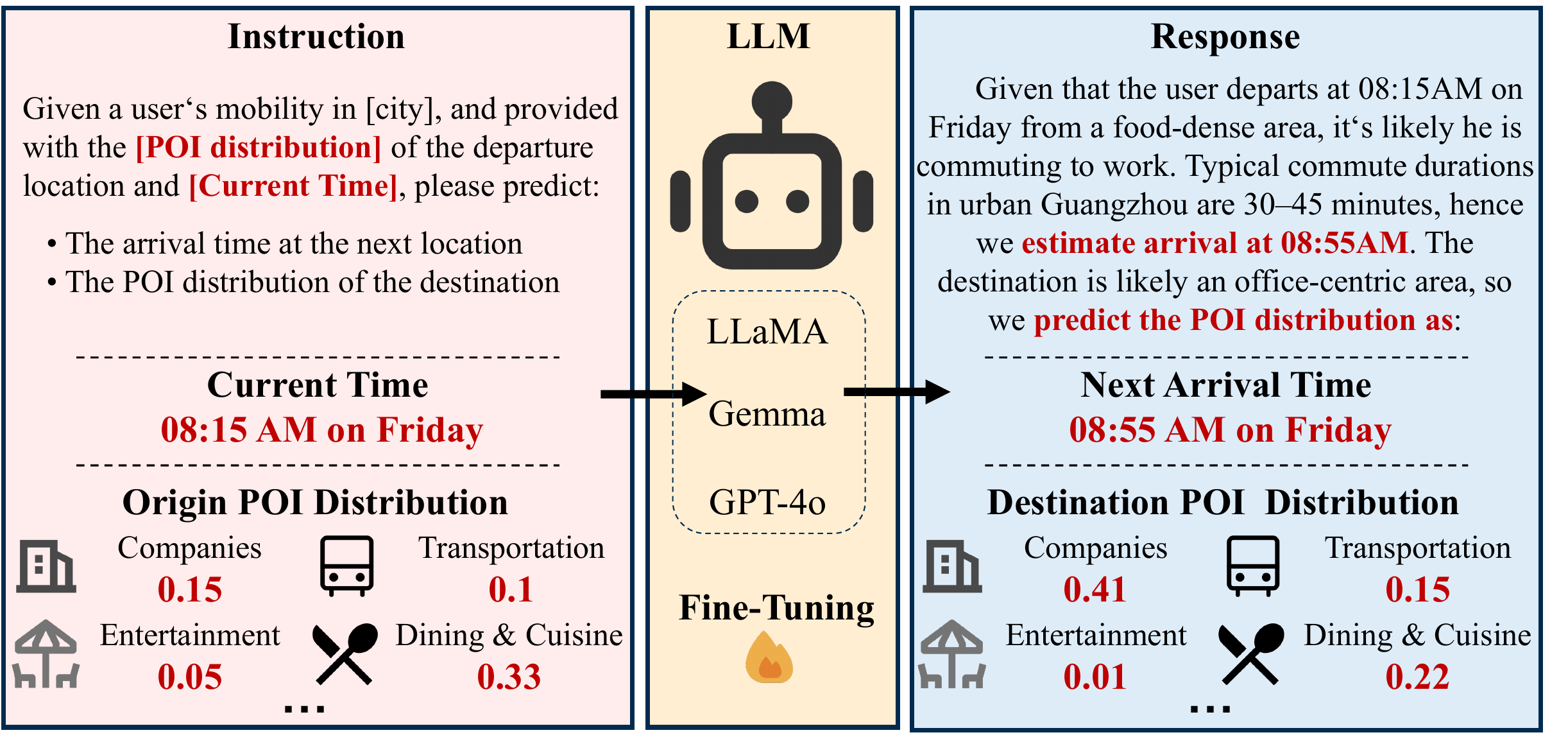}
    \caption{Example of inferring travel plans using fine-tuned LLM. The prompt is structured with task instruction, current time, and origin POI distribution, and the LLM generates temporal and semantic travel plans structured with next arrival time and destination POI distribution.}
    \vspace{-3mm}
    \label{fig:prompt_sample}
\end{figure}

As shown in Figure~\ref{fig:prompt_sample}, at each step, the travel planner reasons over structured contextual inputs, including current time and origin POI distribution, and infers the next arrival time and destination POI distribution. Additionally, we feed inferred next arrival time and destination POI distribution through an output layer to obtain logits results, indicating the probability of each time slot or POI category as the predicted. Mathematically, given a prompt $p$ constructed from current time $t_k$ and regional origin POI distribution $\mathbf{p}_k$, travel planner produces:
\begin{equation}
[\hat{y}_{\text{time}}, \hat{y}_{\text{poi}}] = \text{OutputLayer}(\text{LLM}(p))
\end{equation}
where \( \hat{y}_{\text{time}} \) and \( \hat{y}_{\text{poi}}\) are logits results of inferred next arrival time slots and destination POI distribution, respectively. By recursively inferring and broadcasting $\hat{y}_{time}$ and $\hat{y}_{poi}$, we obtain the temporal travel plan $M$ and semantic travel plan $D$ as conditions in the diffusion process.

To fine-tune the LLM in the travel planner, we design a dual-objective Kullback–Leibler divergence loss (KLDivLoss)~\cite{Kullback1951Information}. The fine-tuning objective is to minimize a weighted sum of two loss components:
\begin{equation}
\mathcal{L}_{\text{time}} = \text{KLDivLoss}(\text{softmax}(\hat{y}_{\text{time}}), y_{\text{time}}),
\end{equation}
\begin{equation}
\mathcal{L}_{\text{poi}} = \text{KLDivLoss}(\text{softmax}(\hat{y}_{\text{poi}}), y_{\text{poi}}),
\end{equation}
\begin{equation}
\mathcal{L}_{\text{total}} = \mathcal{L}_{\text{poi}} + \lambda \cdot \mathcal{L}_{\text{time}},
\end{equation}
where \( y_{\text{time}}\) denotes the discretized ground truth of the next arrival time, \( y_{\text{poi}}\) denotes the ground truth of POI distribution of the destination, and \( \lambda \) is a hyperparameter that balances next arrival time loss $\mathcal{L}_{\text{time}}$ and POI distribution loss $\mathcal{L}_{\text{poi}}$.

\subsection{Unified Spatial Embedding}
A major limitation of unified human mobility generation across cities is embedding diverse spatial structures into a unified representation space to leverage data in all cities. 
To overcome this limitation, we propose a unified spatial embedding, mapping diverse spatial structures across cities into a unified representation space.
The unified spatial embedding is an encoder-decoder symmetrical architecture for encoding intrinsic attributes of spatial structures and regional sequential features into a unified representation space, and decoding unified representations back into specific spatial identifiers. 
In this manner, spatial structures of all cities can be mapped into a unified representation space, where modeling and generating are conducted. 

\begin{figure*}[tb]
    \centering
    \includegraphics[width=1\linewidth]{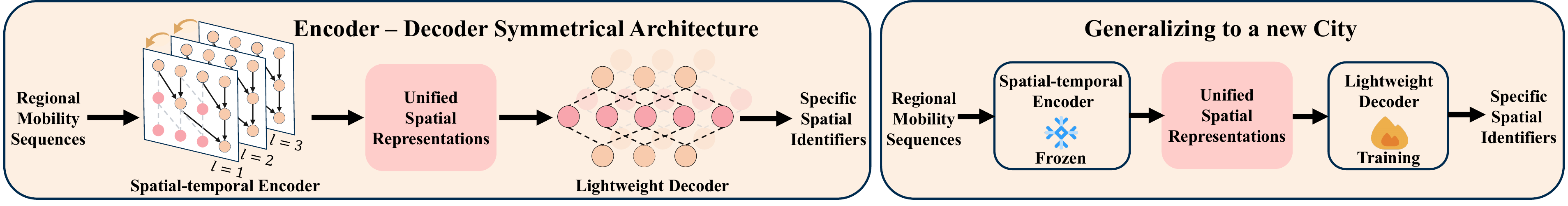}
    \caption{Description of the unified spatial embedding. In this encoder-decoder symmetrical architecture, a spatial-temporal encoder embeds diverse spatial structures in regional mobility sequences into unified spatial representations, and a lightweight decoder maps these representations back onto specific spatial identifiers. When generalizing to a new city, the encoder is frozen and the decoder is trained on a small subset of labeled samples.}
    \label{fig:unified_embedding}
\end{figure*}

We use an encoder to capture spatial and temporal features in a region sequence, embedding spatial structures into a unified representation space. As shown in Figure~\ref{fig:unified_embedding}, the encoder adopts a multi-level temporal convolutional network (TCN) with dilated 1D convolutions and residual paths. The core representation at each layer is computed by fusing information, as formalized by:
\begin{equation}
\Psi^{(l)} = \left(\mathbf{W}_{ker}^{(l)} * \mathbf{E}^{(l)} + \mathbf{b}_{ker}^{(l)}\right) \cdot \delta\left(\mathbf{W}_g^{(l)} * \mathbf{E}^{(l)} + \mathbf{b}_g^{(l)}\right) + \mathbf{E}^{\prime(l)}
\end{equation}
where $E^{(l)}$ is the embedding of POI distribution vector sequences at layer $l$, and ${E'}^{(l)}$ is a residual-enhanced slice for gradient preservation. The convolutional filters $W_{ker}^{(l)}$ and $W_g^{(l)}$, along with biases $b_{ker}^{l}$, $b_g^{l}$, are learned parameters. The sigmoid activation $\delta$ modulates the flow of temporal information. To further enhance the encoding of cross-scale temporal features, we introduce a multi-level correlation injection layer, which incorporates correlations from lower layers into higher ones.

To map unified representations back into specific spatial identifiers, we design a lightweight decoder consisting of a few feed-forward layers. To adapt to an unseen city, we freeze the spatial-temporal encoder and leverage a small amount of labeled data from the unseen city to train the decoder. The lightweight design of the decoder enables the unified spatial embedding to generalize across cities effectively, supporting UniMob to generate human mobility across cities and leverage data from all cities.

\begin{figure}[tb]
    \centering
    \includegraphics[width=0.88\linewidth]{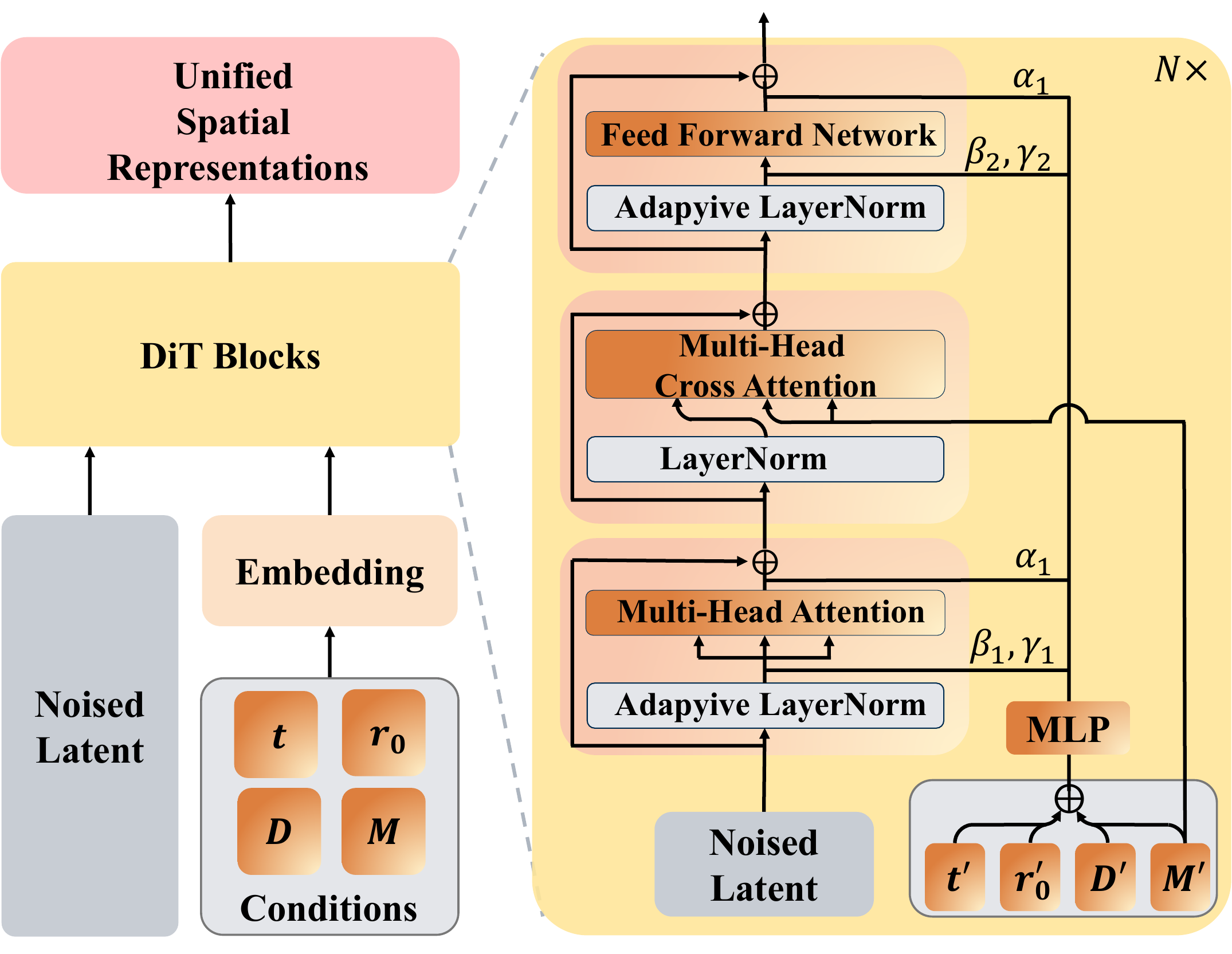}
    \caption{Diffusion Transformer (DiT) Block. Conditions, including time step $t$, start region $r_0$, semantic travel plan $D$ and temporal travel plan $D$, are first embedded into $t'$,$r_0'$,$D'$ and $M'$. By progressively denoising the noised latent, DiT blocks generate unified spatial representations.}
    \label{fig:DiT}
\end{figure}

\subsection{Mobility Generator}
\label{Mobility_Generator}
Human mobility is influenced by complex factors, and the variability and stochasticity are pronounced when modeling across cities. To capture the diversity and generate human mobility across cities, we propose a diffusion mobility generator, using the start region as an anchor location, the temporal travel plans to encode temporal dynamics, semantic travel plans to model travel patterns. By progressively denoising latent vectors, the mobility generator captures both the globally consistent regularities and variability of human mobility across cities, probabilistically generating high-fidelity and individual human mobility. Mathematically, the mobility generator is formulated as:
\begin{equation}
\hat{\mathbf{x}}_0 = f_\theta(\mathbf{x}_t, t, r_0, M, D),  
\end{equation}
where$x_0$ denotes the generated unified mobility representations, $x_t$ denotes the noisy latent representation at diffusion step $t$, $r_0$ denotes start region. $M$ and $D$ are temporal mode and semantic intent guidance obtained from the LLM-based travel planner.

Mobility generator leverages a conditional diffusion transformer (DiT) as the backbone. As shown in Figure~\ref{fig:DiT}, within each block, conditions are first embedded and then directly added. This combined vector is passed through an MLP to produce the modulation parameters. Specifically, the transformation parameters are generated as:
\begin{equation}
    \gamma_1, \beta_1, \alpha_1, \gamma_2, \beta_2, \alpha_2 = \text{MLP}(\mathbf{r}_0' + \mathbf{M}' + D' + t')
\end{equation}
where $r_0'$, $M'$, $D'$, and $t'$ are embedded representations of the start region, temporal mode, semantic intent guidance, and diffusion step respectively. The resulting scale and shift parameters $(\gamma,\beta,\alpha)$ are used to modulate the latent features across layers. Subsequently, the normalized output from AdaLN is refined through a Multi-Head Self-Attention (MHSA) mechanism to capture complex dependencies and enhance the representation. The output $\mathbf{X}_t'$ of this process can be formalized as:
\begin{equation}
    \mathbf{X}'_t = \alpha_1 \cdot \text{MHSA}(\text{AdaLN}(\mathbf{X}_t; \gamma_1, \beta_1)) + \mathbf{X}_t
\end{equation}
To further enhance temporal alignment and control, a MultiHead Cross-Attention (MHCA) mechanism is incorporated into the DiT block, positioned between the MHSA layer and the feed-forward network (FFN). The MHCA module interacts with the condition embedding $M'$ after applying layer normalization (LN), enabling the effective integration of temporal dynamics into the representation. The output $\mathbf{X}_t''$ of this combined process can be formalized as:
\begin{equation}
    \mathbf{X}''_t = \text{MHCA}(\text{LN}(\mathbf{X}'_t), \mathbf{M}', \mathbf{M}') + \mathbf{X}'_t
\end{equation}
The residual connection within the FFN step ensures stability and preserves essential information, leading to the final refined mobility representation $\hat{\mathbf{X}}_0$ as follows:
\begin{equation}
    \hat{\mathbf{X}}_0 = \alpha_2 \cdot \text{FFN}(\text{AdaLN}(\mathbf{X}''_t; \gamma_2, \beta_2)) + \mathbf{X}''_t
\end{equation}
This design enables the DiT Block to effectively guide denoising according to conditions, resulting in the generation of high-quality human mobility.

\begin{table*}[t]
\centering
\caption{Statistics of the two real-world datasets used in our experiments.}
\label{tab:dataset_stats}
\setlength{\tabcolsep}{5pt}
\begin{tabular}{lcccc lcccc}
\toprule
\multicolumn{5}{c}{\textbf{Private Car Dataset}} & \multicolumn{5}{c}{\textbf{Mobile Phone Dataset}} \\
\cmidrule(lr){1-5} \cmidrule(lr){6-10}
\textbf{City} & \textbf{\#Records} & \textbf{\#Vehicles} & \textbf{Time Span} & \textbf{\#POIs} &
\textbf{City} & \textbf{\#Records} & \textbf{\#Users} & \textbf{Time Span} & \textbf{\#POIs} \\
\midrule
Guangzhou & 98280 & 29180 & 2024/01/01--2024/01/15 & 793057 &
Shanghai  & 89966 & 9481 & 2024/06/01--2024/06/30 & 875979 \\
Shenzhen  & 136675 & 17245 & 2022/08/01--2022/08/15 & 685218 &
Beijing   & 105777 & 10000 & 2019/12/01--2019/12/15 & 757969 \\
Changsha  & 62957 & 16578 & 2022/12/01--2022/12/15 & 406490 &
Shenzhen  & 69964 & 10000 & 2021/11/01--2021/11/07 & 685218 \\
\bottomrule
\end{tabular}
\end{table*}

\section{Experiments}
\label{experiments}
In this section, we conduct comprehensive experiments to evaluate the effectiveness, generalizability, interpretability, and privacy implications of UniMob for generating human mobility.

\subsection{Experimental Setup}
\subsubsection{Datasets}
We conduct experiments using two real-world human mobility datasets: 

\textbf{Private car dataset} consists of GPS traces collected from private vehicles in Guangzhou, Shenzhen, and Changsha. To protect user privacy, the original coordinate-level locations are mapped to region-level identifiers using a Voronoi partitioning scheme based on population heatmaps, ensuring that spatial data is aggregated without compromising sensitive information. Each data entry includes a de-identified vehicle ID, timestamp, and corresponding region ID. We select one month of trip data from the three representative cities as the basis for private vehicle mobility modeling.

\textbf{Mobile phone dataset} includes mobile user location traces obtained from two different sources: GPS-based positioning data for users in Beijing and Shenzhen, and cell-tower-based localization records for users in Shanghai. All user locations are anonymized and discretized into region-level identifiers using a similar partitioning strategy as the car dataset. Each data entry contains a de-identified user ID, timestamp, and region ID, providing insights into human mobility patterns.

It is important to emphasize that all data utilized in this study was collected with the explicit authorization of both individual users and the data provider. All data were region-level aggregates that had been processed and anonymized by the provider before our access. 

\begin{table}[h]
    \centering
    \caption{List of POI categories.}
    \vspace{-2mm}
    \label{tab:poi_cate}
    \begin{tabular}{ll}
        \toprule
        \multicolumn{2}{c}{\textbf{Default 14 POI Categories}} \\
        \midrule
        Transportation Facilities & Leisure \& Entertainment \\
        Companies \& Enterprises  & Healthcare \\
        Commercial \& Residential & Tourist Attractions \\
        Automotive                & Life Services \\
        Science \& Education \& Culture & Shopping \& Consumer Goods \\
        Sports \& Fitness         & Hotels \& Accommodations \\
        Financial Institutions    & Dining \& Cuisine \\
        \bottomrule
    \end{tabular}
\end{table}

\begin{table*}[tb]
    \centering
    \vspace{-3mm}
    \caption{Performance of UniMob in human mobility generation on two real-world datasets: private car dataset and mobile phone dataset from major cities in China. The smaller metrics mean better performance, and bold indicates the best method and the underline indicates the second-best method.}
    \vspace{-3mm}
    \label{tab:baseline}
    \resizebox{\textwidth}{!}{
    \begin{tabular}{c ccccccccccccccc} 
    \toprule
   & \multicolumn{15}{c}{\textbf{Private car dataset}} \\
      \cmidrule(lr){1-16}
       \textbf{city}&\multicolumn{5}{c}{\textbf{Guangzhou}}&\multicolumn{5}{c}{\textbf{Shenzhen}}&\multicolumn{5}{c}{\textbf{Changsha}}\\
    \cmidrule(lr){2-6}\cmidrule(lr){7-11}\cmidrule(lr){12-16}
 \textbf{Metrics}& Distance & Radius & LocNum & G-rank & R-rank & Distance & Radius & LocNum & G-rank & R-rank & Distance & Radius & LocNum & G-rank & R-rank\\
    \hline
    DRTISA & 0.1654 & 0.2283 & 0.0655 & 0.0205 & 0.1376 & 0.1945 & 0.2403 & 0.0731 & 0.0588 & 0.0882 & 0.2479 & 0.2111 & 0.0864 & 0.0454 & 0.1567 \\
    DensityEPR & 0.4080 & 0.3338 & 0.3683 & 0.0339 & 0.1232 & 0.3296 & 0.3350 & 0.4148 & 0.0390 & 0.1260 & 0.3589 & 0.3190 & 0.3861 & 0.0384 & 0.1632 \\
    SeqGAN & 0.1246 & 0.1193 & 0.3670 & \textbf{0.0013} & 0.0881 & 0.0948 & 0.2108 & 0.1822 & \underline{0.0019} & \underline{0.0745} & 0.1179 & 0.3730 & 0.3012 & 0.0091 & 0.0986 \\
    MoveSim & 0.1947 & 0.1460 & 0.4007 & 0.0077 & 0.1653 & 0.1720 & 0.1797 & 0.2668 & 0.0182 & 0.1380 & 0.0763 & 0.1947 & \underline{0.0669} & 0.0069 & 0.1212 \\
    DiffTraj & \underline{0.0814} & \underline{0.1149} & \underline{0.0626} & 0.0037 & 0.2963 & \underline{0.0865} & \underline{0.1281} & \underline{0.0612} & 0.0273 & 0.1256 & \underline{0.0654} & \underline{0.0459} & 0.1368 & \underline{0.0032} & 0.2920 \\
    TrajGDM & 0.3284 & 0.2451 & 0.4327 & 0.0028 & 0.1107 & 0.1508 & 0.2849 & 0.2832 & 0.0028 & 0.1107 & 0.1433 & 0.3028 & 0.3836 & 0.0062 & 0.0958 \\
    LLMob & 0.1843 & 0.1269 & 0.2340 & 0.0210 & \underline{0.0562} & 0.1250 & 0.1421 & 0.2403 & 0.0214 & 0.0989 & 0.1192 & 0.1504 & 0.3287 & 0.0339 & \textbf{0.0589} \\
    LLM-COD & 0.1264 & 0.1368 & 0.3060 & 0.0167 & 0.0719 & 0.1768 & 0.1504 & 0.3287 & 0.0330 & 0.1256 & 0.1245 & 0.1287 & 0.2776 & 0.0379 & 0.1537 \\
    \rowcolor{gray!20} \textbf{OURS} & \textbf{0.0672} & \textbf{0.0456} & \textbf{0.0451} & \underline{0.0014} & \textbf{0.0396} & \textbf{0.0650} & \textbf{0.0451} & \textbf{0.0035} & \textbf{0.0018} & \textbf{0.0468} & \textbf{0.0637} & \textbf{0.0284} & \textbf{0.0505} & \textbf{0.0030} & \underline{0.0600} \\
 \midrule
 \midrule
 \addlinespace[2.5pt]
    & \multicolumn{15}{c}{\textbf{Mobile phone dataset}} \\
      \cmidrule(lr){1-16}
 \textbf{city}&\multicolumn{5}{c}{\textbf{Beijing}}&\multicolumn{5}{c}{\textbf{Shenzhen}}&\multicolumn{5}{c}{\textbf{Shanghai}}\\
     \cmidrule(lr){2-6}\cmidrule(lr){7-11}\cmidrule(lr){12-16}
 \textbf{Metrics}& Distance & Radius & LocNum & G-rank & R-rank & Distance & Radius & LocNum & G-rank & R-rank & Distance & Radius & LocNum & G-rank & R-rank\\
    \hline
    DRTISA & 0.0264 & 0.2507 & 0.2146 & 0.0073 & 0.0185 & 0.0218 & 0.2360 & 0.4440 & 0.0032 & 0.1323 & 0.0473 & 0.2240 & 0.3210 & 0.0318 & 0.0338 \\
    DensityEPR & 0.0367 & 0.4088 & 0.4075 & 0.1961 & 0.2992 & 0.0209 & 0.1492 & 0.2133 & \underline{0.0019} & \underline{0.0173} & 0.0262 & 0.4216 & 0.4790 & 0.0870 & 0.0647 \\
    SeqGAN & 0.0822 & 0.3949 & 0.3657 & 0.0832 & 0.1239 & \underline{0.0169} & 0.2086 & 0.2301 & 0.0739 & 0.0354 & 0.0348 & 0.2152 & 0.4800 & 0.0376 & 0.0587 \\
    MoveSim & 0.0861 & 0.3245 & 0.2900 & 0.0197 & 0.1074 & 0.0234 & 0.2206 & 0.2555 & 0.0165 & 0.0507 & 0.0262 & 0.2313 & 0.4289 & 0.0561 & 0.0964 \\
    DiffTraj & \textbf{0.0059} & \underline{0.0069} & \underline{0.0188} & 0.0072 & \underline{0.0053} & 0.0301 & \underline{0.0413} & \underline{0.0829} & 0.2268 & 0.0251 & \textbf{0.0166} & \underline{0.0519} & \underline{0.1617} & \underline{0.0073} & \underline{0.0175} \\
    TrajGDM & 0.0479 & 0.2714 & 0.2685 & \underline{0.0068} & 0.0607 & 0.0295 & 0.2489 & 0.2236 & 0.0091 & 0.0630 & 0.0239 & 0.2977 & 0.4121 & 0.1028 & 0.0491 \\
    LLMob & 0.0456 & 0.1333 & 0.1493 & 0.0234 & 0.0719 & 0.0225 & 0.2417 & 0.1234 & 0.1185 & 0.0886 & 0.0386 & 0.2276 & 0.2706 & 0.0244 & 0.0833 \\
    LLM-COD & 0.0765 & 0.3441 & 0.2918 & 0.0180 & 0.0989 & 0.0591 & 0.1787 & 0.2072 & 0.1742 & 0.0714 & 0.0211 & 0.2162 & 0.2544 & 0.0134 & 0.0878 \\
    \rowcolor{gray!20} \textbf{OURS} & \underline{0.0077} & \textbf{0.0050} & \textbf{0.0116} & \textbf{0.0060} & \textbf{0.0025} & \textbf{0.0097} & \textbf{0.0141} & \textbf{0.0187} & \textbf{0.0013} & \textbf{0.0072} & \underline{0.0180} & \textbf{0.0173} & \textbf{0.0404} & \textbf{0.0018} & \textbf{0.0169} \\
    \bottomrule
    \end{tabular}
    }
    \vspace{-3mm}
\end{table*}

To capture the functional semantics of each region, we construct a POI profile by aggregating POIs. We obtain POI data from publicly available map APIs, which include geographic coordinates, category labels, and place names. To facilitate regional semantic modeling, we map each POI to its corresponding spatial region (either Voronoi cell or 1km grid cell) and aggregate POIs within each region. Each POI is assigned to one of 14 high-level semantic categories based on its original label. The POI list is shown in Table~\ref{tab:poi_cate}. For each region, we compute the proportion of POIs falling into each category, resulting in a 14-dimensional POI distribution vector. These vectors are used to represent region-level semantics in our framework. In this way, each sample comprises a start timestamp, start region, a sequence of region IDs denoting mobility, and associated POI distributions.

\subsubsection{Metrics}
We use five metrics to assess the quality of the generated data, illustrating the spatiotemporal characteristics of mobility, involving \emph{Distance}, \emph{Radius}, \emph{LocNum}, \emph{G-rank}, \emph{R-rank}. 

\begin{itemize}
    \item \textbf{Distance}: Measures the average travel distance for each movement segment within mobility.
    \item \textbf{Radius}: Captures the radius of gyration, indicating how far individuals typically move from their mobility centroid.
    \item \textbf{LocNum}: Counts the number of unique locations visited by an individual.
    \item \textbf{G-rank}: Represents the global visit frequency distribution over the top 100 most visited regions across all users.
    \item \textbf{R-rank}: Aggregates destination preferences by ranking the top 50 most frequently visited regions across all users who start from the same origin region.
\end{itemize}

The Jensen-Shannon Divergence (JSD)~\cite{Menendez1997JSD} is utilized to quantify the similarity between the distributions of generated and real human mobility under each metric. We adopt the Jensen–Shannon (JS) divergence. It is a symmetric, smoothed variant of the Kullback–Leibler (KL) divergence and is defined as:

\begin{equation}
D_{JS}(\mathbf{p}, \mathbf{q}) = \frac{1}{2} D_{KL} \left( \mathbf{p} \middle\| \frac{\mathbf{p} + \mathbf{q}}{2} \right) + \frac{1}{2} D_{KL} \left( \mathbf{q} \middle\| \frac{\mathbf{p} + \mathbf{q}}{2} \right),
\end{equation}

\noindent where $D_{KL}(p \| q)$ is KL divergence, equal to $\sum_{x \in X} p(x) \log \left( \frac{p(x)}{q(x)} \right)$, $p$ and $q$ are two distributions, respectively. The KL divergence is given by:
\begin{equation}
D_{KL}(\mathbf{p} \| \mathbf{q}) = \sum_{x \in X} \mathbf{p}(x) \log \left( \frac{\mathbf{p}(x)}{\mathbf{q}(x)} \right).
\end{equation}

\noindent The JS divergence ranges from 0 to 1, where lower values indicate greater similarity between the generated and real distributions, suggesting higher fidelity of the synthetic mobility.

\subsubsection{Baselines}
To evaluate the performance of our method, we compare it with the following eight state-of-the-art models:

$\bullet$ \textbf{DRTISA} (Data-Driven Region Transition and Individual Simulation Algorithm). DRTISA~\cite{pappalardo2018data} is a rule-based simulator that models individual movements by learning region transition probabilities from data. It simulates trajectories by sampling from learned distributions while enforcing user-level constraints such as departure frequency and preferred destinations.

$\bullet$ \textbf{DensityEPR}. DensityEPR~\cite{pappalardo2016human} extends the classic Exploration and Preferential Return (EPR) framework with regional population density information. It models human mobility as a balance between exploration of new regions and return to previously visited ones, using a probabilistic mechanism grounded in spatial density distributions.

$\bullet$ \textbf{SeqGAN}. SeqGAN~\cite{yu2017seqgan} applies Generative Adversarial Networks to sequential data, treating trajectory generation as a reinforcement learning problem. A generator produces plausible region sequences while a discriminator distinguishes generated trajectories from real ones, guiding the generator through adversarial feedback.

$\bullet$ \textbf{MoveSim}. MoveSim~\cite{feng2020learning} is a GAN-based human mobility simulator that conditions trajectory generation on user identity and historical movement. It captures semantic similarity across trajectories by learning latent representations, enabling generation that reflects individual preferences and spatial patterns.

$\bullet$ \textbf{DiffTraj}. DiffTraj~\cite{zhu2023difftraj} introduces denoising diffusion probabilistic models into trajectory generation. It transforms observed trajectories into a continuous latent space, and generates new samples through iterative denoising, allowing for multimodal trajectory prediction with uncertainty modeling.

$\bullet$ \textbf{TrajGDM}. TrajGDM~\cite{chu2024simulating} extends diffusion modeling with guidance modules that encode spatial and temporal conditions. By injecting control signals into the reverse diffusion process, it allows the model to generate trajectories that better align with semantic and environmental constraints.

$\bullet$ \textbf{LLMob}. LLMob~\cite{jiawei2024large} utilizes a language model to model and predict human movement through instruction-tuned generation. Given natural language prompts describing context (e.g., time, location type), it generates next-location predictions via autoregressive decoding.

$\bullet$ \textbf{LLM-COD}. LLM-COD~\cite{yu2024harnessing} formulates destination prediction as a conditional language modeling task. It uses prompts that encode departure location, time, and regional POI context to guide the LLM in generating plausible destination semantics or region IDs.

\begin{figure}[h]
    \centering
    \vspace{-3mm}
    \includegraphics[width=0.9\linewidth]{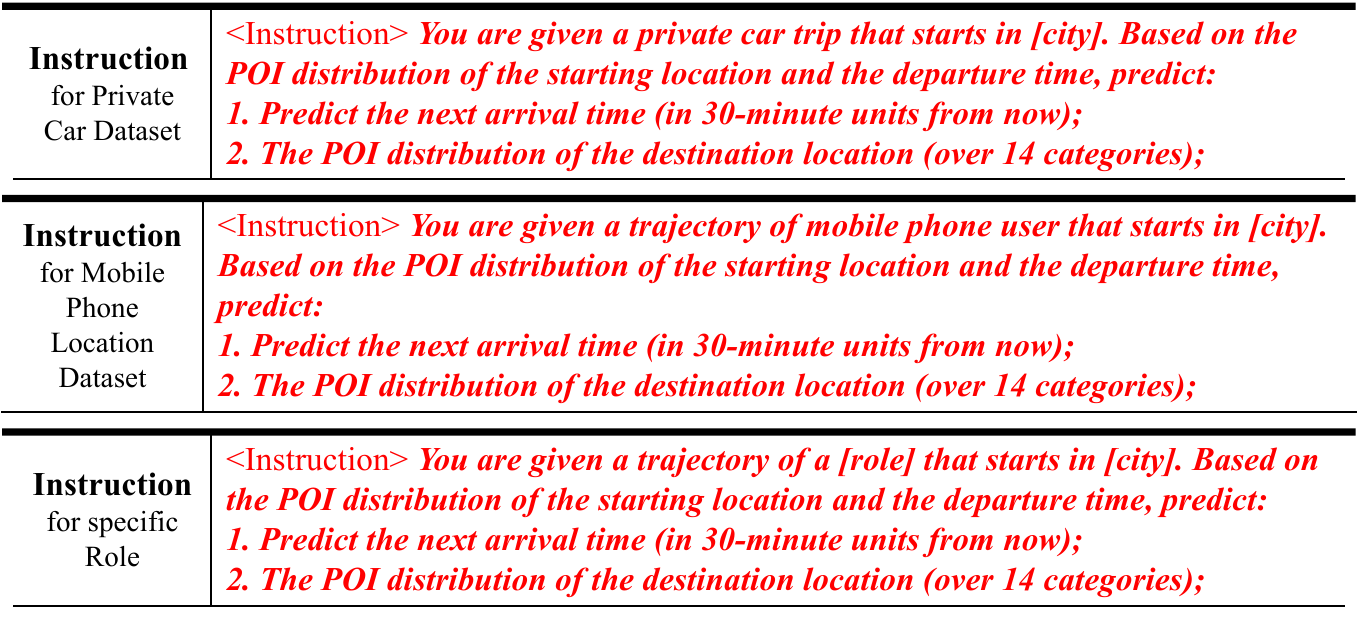}
    \caption{Instruction templates for different datasets and user role profiles.}
    \vspace{-3mm}
    \label{fig:appendix_instruction}
    \vspace{-3mm}
\end{figure}

\subsubsection{Experimental Settings}
We design different prompt templates for different datasets and user role profiles to guide the LLM in generating semantically meaningful mobility. Instructions illustrated in Figure~\ref{fig:appendix_instruction} are used as input during travel planning to provide contextual cues.
We adopt a multi-city training strategy for UniMob, where training samples for each dataset are drawn from multiple cities. The travel planner, the mobility generator, and the spatial-temporal encoder are trained on the combined data. In contrast, the lightweight decoder is trained using a small portion of data from a specific city. Specifically, we randomly sample 50,000 samples for each city in both datasets and split them into 80\% for training, 10\% for validation, and 10\% for testing. In few-shot settings, only 10\% of the target city's training samples are used, while the lightweight decoder is tuned with 5\% of the amount of the data used for multi-city training.

Unless otherwise specified, UniMob is configured with a latent dimension of 128, a feedforward dimension of 2048, four DiT blocks, and eight attention heads. The diffusion process runs for 1000 steps with a beta schedule linearly increasing from 0.001 to 0.1. The spatial-temporal encoder employs 1D convolutions with kernel size 3, hidden dimension 32, and dilation factor 1. All experiments are conducted on a single NVIDIA A100 - 40GB PCIe GPU, using Xavier initialization for parameters and the Adam optimizer for training.

\subsection{Overall Performance}
Extensive experiments on two real-world datasets consistently demonstrate that UniMob outperforms state-of-the-art baselines by significant margins across all evaluation metrics. As shown in Table~\ref{tab:baseline}, UniMob achieves remarkable improvements in critical metrics, which quantify the spatial characteristics and movement patterns of generated mobility. For instance, in the private car dataset for Guangzhou, UniMob yields a Distance metric of 0.0672, outperforming DiffTraj (0.0814) and MoveSim (0.1947) by 17.4\% and 65.5\% respectively. Similarly, in the mobile phone dataset for Beijing, UniMob achieves a Radius score of 0.0050, which is substantially lower than LLMob (0.1333) and TrajGDM (0.2714), indicating superior alignment with real-world human mobility.

The consistent outperformance across all cities in both datasets underscores the effectiveness of UniMob's unified design. The model's ability to maintain strong performance across diverse urban environments validates its core objective of providing equitable support for mobility generation regardless of a city's size or data resources. Notably, the superior performance can be attributed to the synergistic effect of its three components: the LLM-based travel planner ensures semantically meaningful mobility intent, the unified spatial embedding enables cross-city generalization, and the diffusion-based mobility generator captures fine-grained spatiotemporal dynamics .

Further analysis of the results reveals that UniMob particularly excels in preserving the intricate balance between global regularities and local variations in human mobility. This is evident in the low scores across G-rank and R-rank metrics, which measure the alignment of visit frequency distributions with real data at both global and origin-specific levels. Such performance confirms that UniMob generates mobility not only statistically similar to real mobility but also retains the context-aware patterns critical for downstream applications.

\begin{table}[h]
\centering
\caption{Performance of UniMob and a baseline model (MoveSim) in human mobility generation under limited supervision: 100\%, 75\%, 50\%, and 25\% of the default multi-city training data. The smaller metrics mean better performance.}
\label{tab:reduced_training_size}
\vspace{-2mm}
\begin{tabular}{lccccc}
\toprule
\multicolumn{6}{c}{\textbf{Private car dataset (Guangzhou)}} \\
\midrule
\textbf{Data Size} & Distance & Radius & LocNum & G-rank & R-Rank \\
\midrule
100\%    & 0.0672 & 0.0456 & 0.0451 & 0.0014 & 0.0396 \\
75\%     & 0.0850 & 0.0852 & 0.0666 & 0.0015 & 0.0462 \\
50\%     & 0.0964 & 0.1007 & 0.0859 & 0.0029 & 0.0960 \\
25\%     & 0.1023 & 0.1171 & 0.1153 & 0.0036 & 0.1224 \\
MoveSim  & 0.1947 & 0.1460 & 0.4007 & 0.0077 & 0.1653 \\
\midrule
\multicolumn{6}{c}{\textbf{Mobile phone dataset (Beijing)}} \\
\midrule
\textbf{Data Size} & Distance & Radius & LocNum & G-rank & R-Rank \\
\midrule
100\%    & 0.0077 & 0.0050 & 0.0116 & 0.0060 & 0.0025 \\
75\%     & 0.0083 & 0.0081 & 0.0138 & 0.0078 & 0.0085 \\
50\%     & 0.0120 & 0.0136 & 0.0150 & 0.0122 & 0.0198 \\
25\%     & 0.0289 & 0.0418 & 0.0184 & 0.0154 & 0.0359 \\
MoveSim  & 0.0861 & 0.3245 & 0.2900 & 0.0197 & 0.1074 \\
\bottomrule
\end{tabular}
\vspace{-2mm}
\end{table}

\begin{table*}[tb]
    \centering
    \vspace{-3mm}
    \caption{Performance of UniMob in human mobility generation in unseen cities under different supervision settings: zero-shot, few-shot, and full supervision. The smaller metrics mean better performance.}
    \vspace{-3mm}
    \label{tab:generalization}
    \small 
    \setlength{\tabcolsep}{4pt} 
    \begin{tabular}{c ccccccccccccc} 
    \toprule
    & \multicolumn{12}{c}{\textbf{Private car dataset}} \\
    \cmidrule(lr){1-13}
    \textbf{Setting} & \multicolumn{4}{c}{\textbf{Guangzhou, Shenzhen$\Rightarrow$Changsha}} & \multicolumn{4}{c}{\textbf{Changsha, Guangzhou$\Rightarrow$Shenzhen}} & \multicolumn{4}{c}{\textbf{Changsha, Shenzhen$\Rightarrow$Guangzhou}} \\
    \cmidrule(lr){2-5} \cmidrule(lr){6-9} \cmidrule(lr){10-13}
    \textbf{Metrics} & Distance & Radius & LocNum & G-rank & Distance & Radius & LocNum & G-rank & Distance & Radius & LocNum & G-rank \\
    \midrule
    Zero & 0.0726 & 0.0305 & 0.0513 & 0.0032 & 0.0788 & 0.0607 & 0.0042 & 0.0047 & 0.0736 & 0.0578 & 0.0509 & 0.0073 \\
    Few & 0.0676 & 0.0287 & 0.0511 & 0.0030 & 0.0709 & 0.0605 & 0.0041 & 0.0037 & 0.0713 & 0.0537 & 0.0453 & 0.0066 \\
    \rowcolor{gray!10} Full & \textbf{0.0637} & \textbf{0.0284} & \textbf{0.0505} & \textbf{0.0030} & \textbf{0.0650} & \textbf{0.0451} & \textbf{0.0035} & \textbf{0.0018} & \textbf{0.0672} & \textbf{0.0456} & \textbf{0.0451} & \textbf{0.0014} \\
    \midrule
    \midrule
    \addlinespace[2.5pt]
    & \multicolumn{12}{c}{\textbf{Mobile phone dataset}} \\
    \cmidrule(lr){1-13}
    \textbf{Setting} & \multicolumn{4}{c}{\textbf{Shanghai, Shenzhen$\Rightarrow$Beijing}} & \multicolumn{4}{c}{\textbf{Shanghai, Beijing$\Rightarrow$Shenzhen}} & \multicolumn{4}{c}{\textbf{Beijing, Shenzhen$\Rightarrow$Shanghai}} \\
    \cmidrule(lr){2-5} \cmidrule(lr){6-9} \cmidrule(lr){10-13}
    \textbf{Metrics} & Distance & Radius & LocNum & G-rank & Distance & Radius & LocNum & G-rank & Distance & Radius & LocNum & G-rank \\
    \midrule
    Zero & 0.0379 & 0.0390 & 0.0386 & 0.0234 & 0.0139 & 0.0302 & 0.0256 & 0.0025 & 0.0407 & 0.0305 & 0.0477 & 0.0026 \\
    Few & 0.0091 & 0.0069 & 0.0168 & 0.0185 & 0.1181 & 0.0253 & 0.0221 & 0.0019 & 0.0284 & 0.0259 & 0.0441 & 0.0022 \\
    \rowcolor{gray!10} Full & \textbf{0.0077} & \textbf{0.0050} & \textbf{0.0116} & \textbf{0.0060} & \textbf{0.0097} & \textbf{0.0141} & \textbf{0.0187} & \textbf{0.0013} & \textbf{0.0180} & \textbf{0.0173} & \textbf{0.0404} & \textbf{0.0018} \\
    \bottomrule
    \end{tabular}
    \vspace{-3mm}
\end{table*}

\begin{figure*}[tb]
    \centering
    \vspace{-3mm}
    \subfloat{ \includegraphics[width=0.24\linewidth]{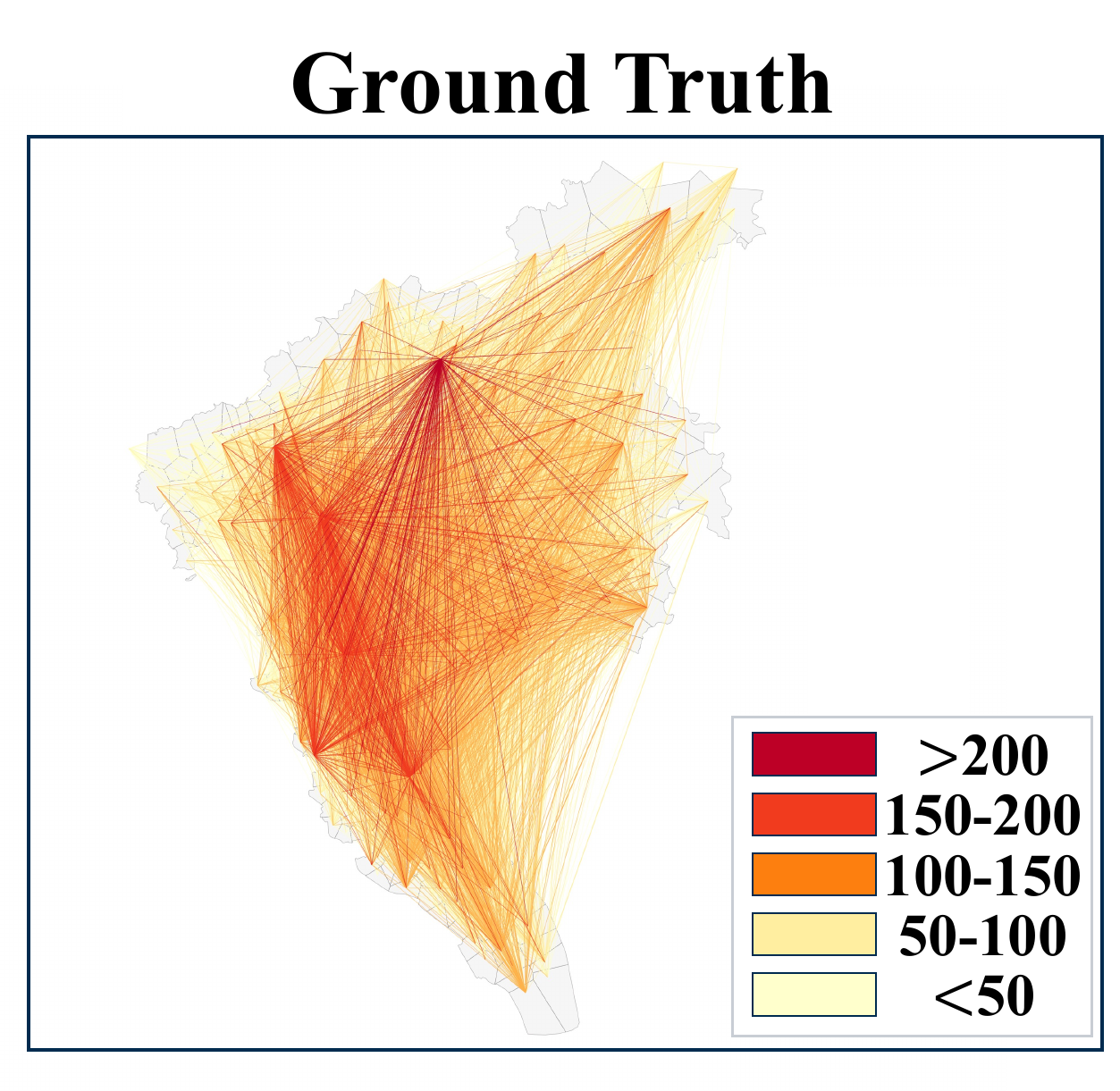}}  
        \subfloat{ \includegraphics[width=0.24\linewidth]{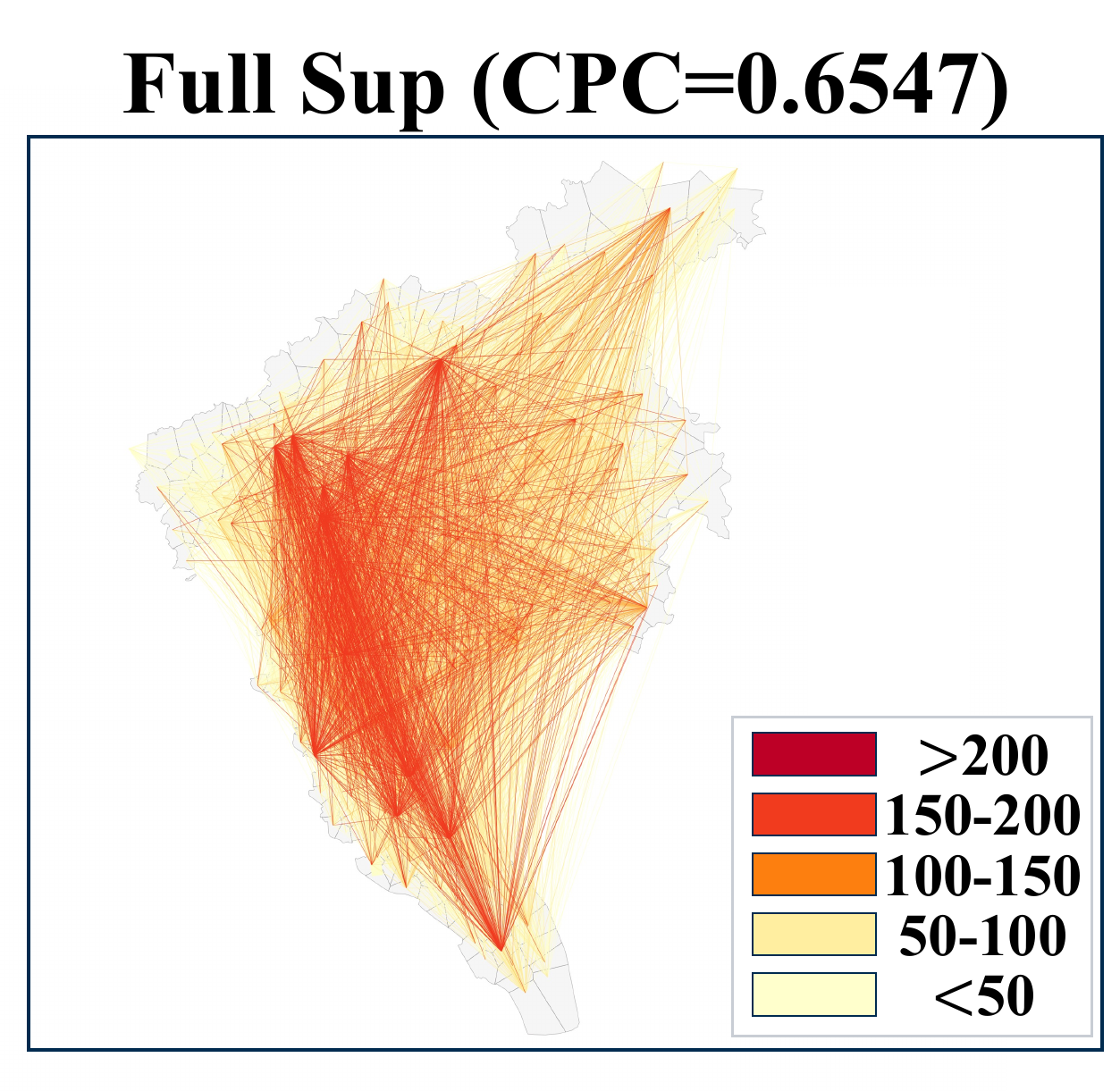}}  
            \subfloat{ \includegraphics[width=0.24\linewidth]{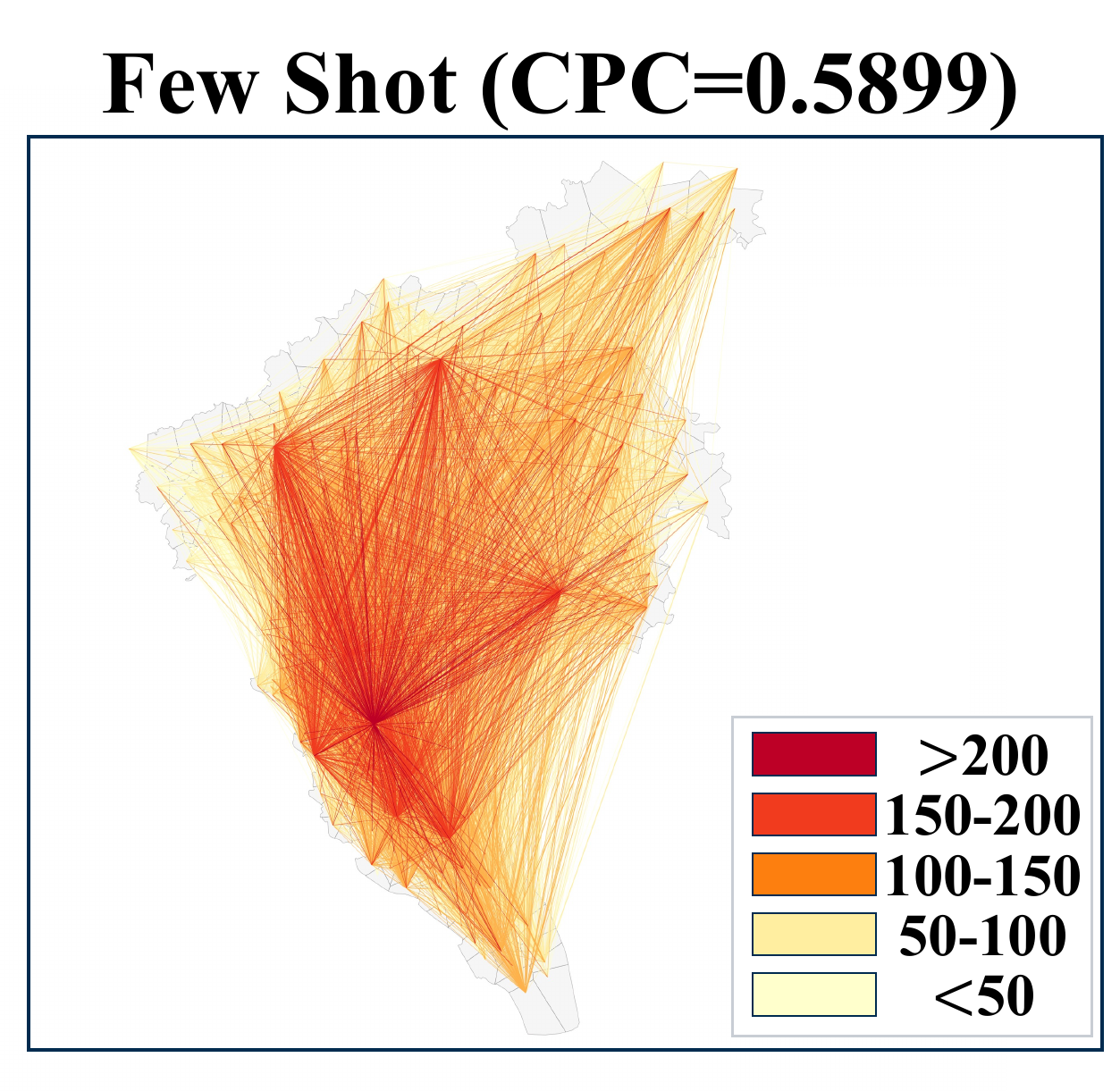}} 
                \subfloat{ \includegraphics[width=0.24\linewidth]{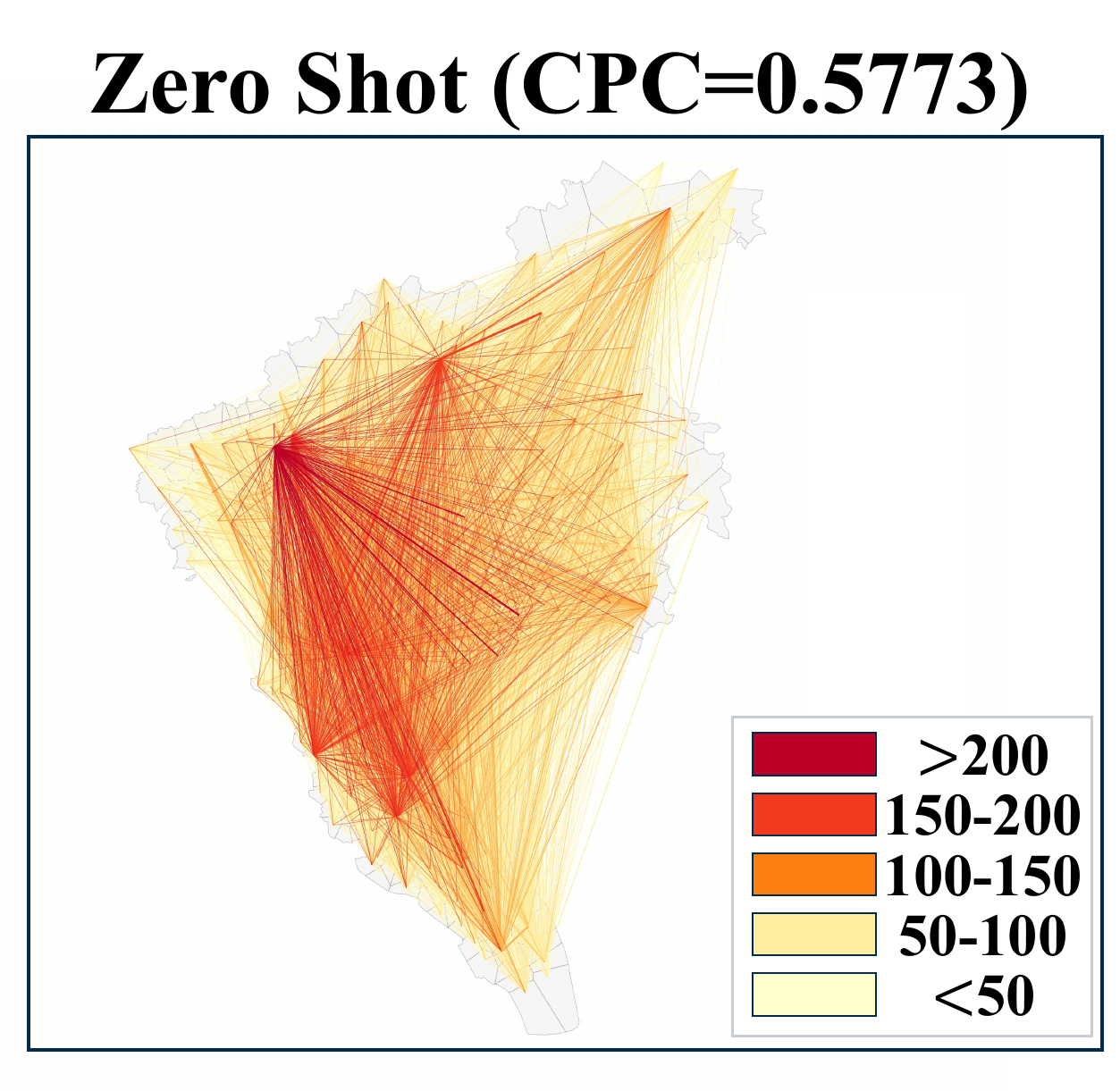}}  
\vspace{-3mm}
    \caption{Visualization of region-level origin-destination (OD) flows generated under Full supervision, Few-shot, and Zero-shot settings, compared to the Ground Truth. The Common Part of Commuters (CPC) metric is used to measure flow similarity.}
    \label{fig:shot}
\vspace{-3mm}
\end{figure*}

\begin{table*}[t]
\centering
\caption{Results of performance of UniMob with different LLM backbones, and performance of ablation variants in human mobility generation. Lower values indicate better performance on all metrics.}
\vspace{-3mm}
\label{tab:llm_ablation}
\footnotesize
\setlength{\tabcolsep}{4.5pt}  
\begin{tabular}{lccccc|ccccc}
\toprule
\multirow{2}{*}{\textbf{Model}} & \multicolumn{5}{c|}{\textbf{Private car dataset (Guangzhou)}} & \multicolumn{5}{c}{\textbf{Mobile phone dataset (Beijing)}} \\
& Distance & Radius & LocNum & G-rank & R-Rank & Distance & Radius & LocNum & G-rank & R-Rank \\
\midrule
OURS-LLaMA2-13B           & 0.0583 & 0.0413 & 0.0448 & 0.0011  & 0.0300  & 0.0071 & 0.0048 & 0.0105 & 0.0056 & 0.0016 \\
OURS-Gemma-2B             & 0.0937 & 0.0650 & 0.0456 & 0.0183  & 0.0588  & 0.0092 & 0.0051 & 0.0140 & 0.0074 & 0.0099 \\
\midrule
w/o Temporal Guiding    & 0.0897 & 0.0588 & 0.0122 & 0.0025  & 0.0419  & 0.0164 & 0.0077 & 0.0707 & 0.0087 & 0.0288 \\
w/o Semantic Guiding & 0.1260 & 0.1703 & 0.1366 & 0.0018  & 0.0468  & 0.0304 & 0.1708 & 0.0187 & 0.0527 & 0.0766 \\
w/o Travel planner        & 0.1521 & 0.1910 & 0.1804 & 0.0426  & 0.0913  & 0.0341 & 0.2623 & 0.0155 & 0.0662 & 0.1216 \\
\rowcolor{gray!15}
OURS-LLaMA2-7B & 0.0673 & 0.0457 & 0.0452 & 0.0014  & 0.0396  & 0.0077 & 0.0050 & 0.0116 & 0.0060 & 0.0025 \\
\bottomrule
\end{tabular}
\vspace{-3mm}
\end{table*}

\subsection{Unified Generation Across Cities}
To assess the capability of UniMob in unified human mobility generation across cities, we first examine how reducing the volume of training data from multiple cities affects mobility synthesis quality. Then we evaluate UniMob’s ability to support mobility generation in unseen cities.

To evaluate whether UniMob can generate high-quality mobility with minimal city-specific supervision by leveraging unified cross-city representations, we reduce the training data size and assess changes in mobility generation performance. As shown in Table~\ref{tab:reduced_training_size}, UniMob maintains strong performance even as the volume of training data decreases. Even with only 25\% of the original data, the model still outperforms the baseline, highlighting its capacity to synthesize high-quality mobility across cities without requiring large amounts of city-specific data.

To assess UniMob's ability to generate high-fidelity human mobility in unseen cities under limited supervision, we design two settings: few-shot and zero-shot. In the few-shot setting, only 10\% of the unseen city's data is used for training. In the zero-shot setting, no unseen city data is used, except for training the lightweight decoder to ensure space alignment.
As shown in~\autoref{tab:generalization}, UniMob consistently performs well across all cities under both few-shot and zero-shot settings. Even without full supervision, UniMob often outperforms baselines trained on completely unseen city data. Notably, transitioning from zero-shot to few-shot brings clear improvements with only minimal additional data, demonstrating UniMob’s practicality in low-resource scenarios.

To further assess the spatial consistency of the generated mobility in unseen cities, we visualize the major origin-destination (OD) flows at the region level under different supervision settings: full, few-shot, and zero-shot. Specifically, we aggregate individual mobility into OD matrices and compare the dominant travel flows in each setting against those observed in the ground truth. As shown in Figure~\ref{fig:shot}, the visualization reveals that UniMob effectively reproduces the spatial structure of real-world mobility, capturing key OD patterns with high fidelity. To quantify this alignment, we compute the Common Part of Commuters (CPC)~\cite{Luca2021SurveyDLMobility} between generated and real OD matrices. The CPC scores confirm the ability of UniMob to generate high-fidelity human mobility even in unseen cities.

\subsection{Role of LLM Guidance}
To evaluate the impact of the travel plan, we first compare different LLM backbones with varying sizes and architectures to assess their effect on mobility generation. We then perform ablation studies to examine the roles of individual components within the travel plan.

{\textbf{LLM Backbone Comparison.}}
To evaluate how LLM architectures influence travel plan inference and mobility generation, we compared three backbones: LLaMA2-13B, LLaMA2-7B, and Gemma-2B, differing in parameter scale and capabilities. Results in Table~\ref{tab:llm_ablation} show LLaMA2-13B outperforms others across metrics, with a 0.0583 Distance in private car dataset (Guangzhou)—13.2\% better than LLaMA2-7B and 37.9\% than Gemma-2B. In mobile phone dataset (Beijing), its 0.0048 Radius exceeds LLaMA2-7B (0.0050) and Gemma-2B (0.0051). Larger models’ superiority stems from stronger semantic reasoning, critical for interpreting contexts. Gemma-2B, though smaller, performs adequately in LocNum/G-rank, suiting resource-limited scenarios but struggling with complex spatiotemporal tasks. This highlights LLM capacity’s role in capturing mobility intent, balancing accuracy and efficiency.

\begin{figure*}[tb]
    \centering
    \includegraphics[width=0.78\linewidth]{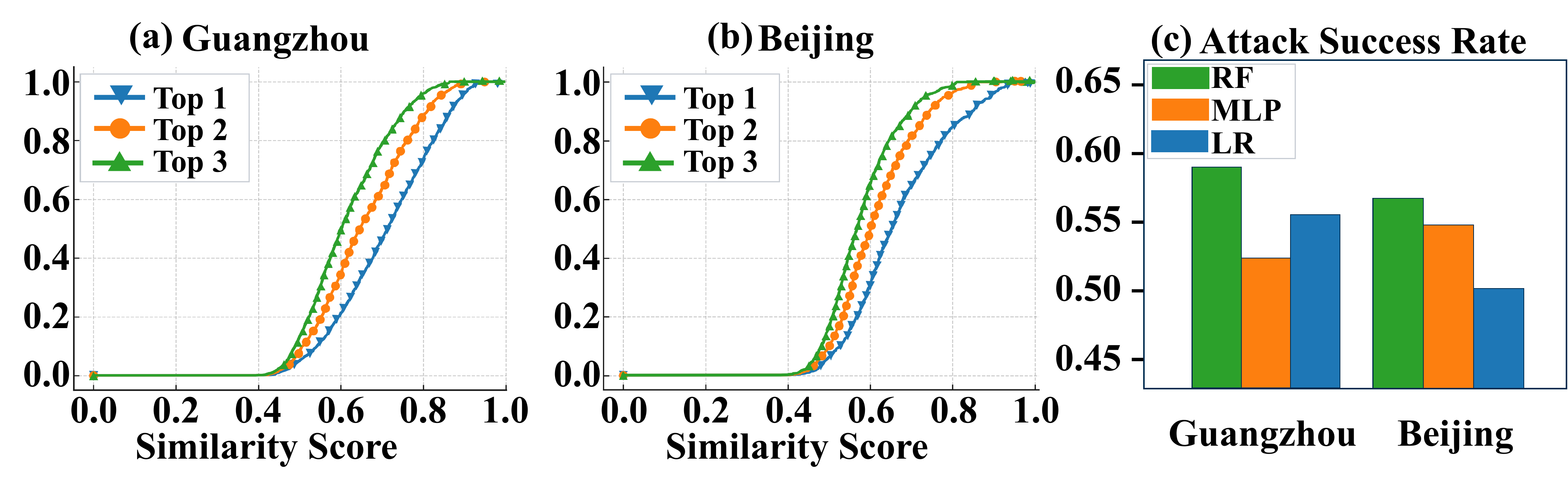}
    \vspace{-3mm}
    \caption{(a) and (b) illustrate results of uniqueness testing on the private car dataset (Guangzhou) and mobile phone dataset (Beijing). For each generated mobility, we compute the similarity scores with its top-1, top-3, and top-5 most similar real counterparts, and analyze the score distribution. (c) demonstrates the success ratios of membership inference attack using different classification algorithms on private car dataset (Guangzhou) and mobile phone dataset (Beijing).}
    \vspace{-3mm}
    \label{fig:privacy}
\end{figure*}

{\textbf{Ablation Study.}}
To examine the specific contributions of each component within the travel planner, we conducted systematic ablation experiments by removing key elements and evaluating the resultant impact on mobility generation performance. Three ablation variants were designed: (1) \textbf{w/o Temporal Travel Plans}, which omits the temporal travel plans inferred by the LLM-based travel planner; (2) \textbf{w/o Semantic Travel Plans}, which excludes the semantic travel plans related to destination POI distributions; and (3) \textbf{w/o Travel Planner}, which entirely removes the travel planner component, eliminating both temporal and semantic guidance for the diffusion-based mobility generator.

Results in Table~\ref{tab:llm_ablation} reveal that removing either temporal or semantic travel plans leads to notable performance degradation across critical metrics, underscoring their distinct but complementary roles. The w/o Temporal Travel Plans variant exhibits significant increases in LocNum (0.0122 vs. 0.0451 in the private car dataset (Guangzhou)) and R-Rank (0.0419 vs. 0.0396), indicating that without temporal guiding, the model struggles to replicate realistic movement frequencies and destination preferences over time. Conversely, the w/o Semantic Travel Plans variant shows marked declines in Distance (0.1260 vs. 0.0672) and Radius (0.1703 vs. 0.0456), reflecting impaired spatial realism when destination POI semantics are ignored.

The most severe performance drop occurs in the w/o Travel Planner variant, which underperforms across all metrics. This comprehensive degradation highlights that temporal and semantic travel plans collectively shape the model’s ability to capture both global regularities and local variations. Together, these findings emphasize that the synergistic integration of temporal and semantic guidance is indispensable for generating high-fidelity human mobility across diverse urban contexts.

\begin{figure}[h]
    \centering
    \includegraphics[width=0.86\linewidth]{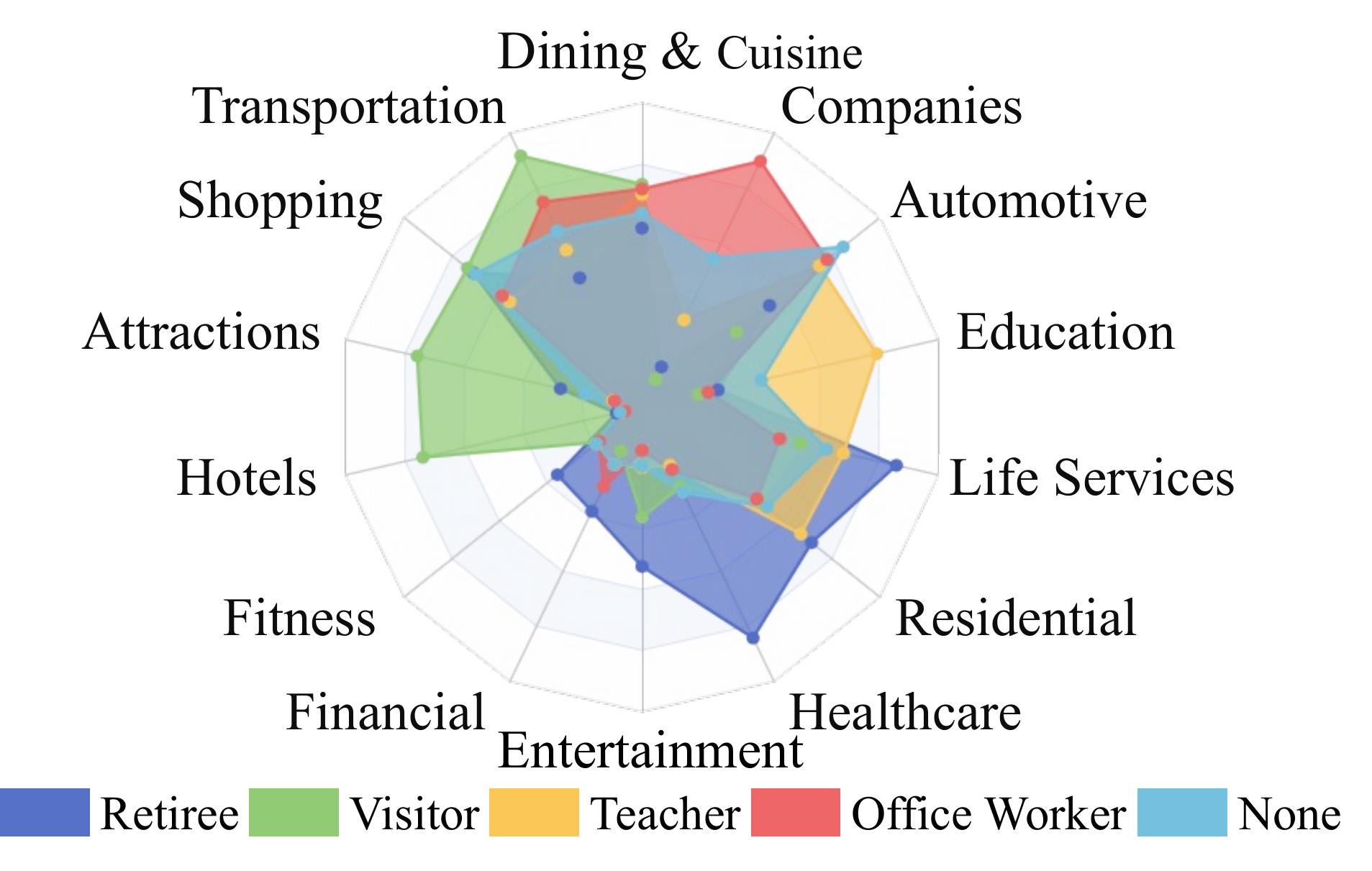}
    \caption{Regional POI composition preference of human mobility generated with four representative user role profiles: retiree, visitor, teacher, and office worker, compared with the results with none profiles.}
    \label{fig:role_radar}
    \vspace{-3mm}
\end{figure}

{\textbf{Impact of Role-specific Instructions.}}
To explore UniMob’s ability to generate personalized and context-aware mobility patterns, we incorporate role-specific instructions into the travel planner, enabling controlled generation tailored to diverse user profiles. We define representative roles and construct corresponding natural language prompts that embed role-specific behavioral traits (e.g., "You are a teacher starting from a residential area at 7:30 AM"). These prompts guide the LLM-based travel planner to infer travel plans aligned with the typical activity rhythms and destination preferences of each role.
Results reveal that role-specific instructions induce distinct semantic shifts in the POI distributions of generated mobility. As visualized in Figure~\ref{fig:role_radar}, mobility generated for "office workers" exhibits a strong preference for regions with high proportions of "Companies \& Enterprises" (41\%) and "Transportation" facilities (15\%), reflecting commuting patterns. In contrast, "retirees" show higher visitation rates to "Dining \& Cuisine" (33\%) and "Life Services" (22\%) regions, aligning with daily errands and leisure activities. This controllability underscores UniMob’s flexibility in simulating population heterogeneity, a critical feature for applications like urban planning and transportation management.

\subsection{Privacy and Utility}
The use of real human mobility data is often restricted due to privacy concerns. In this section, we evaluate whether synthetic mobility poses privacy risks and assess its utility in downstream prediction tasks. To evaluate the privacy implications of the generated mobility, we employ two key analyses: uniqueness testing and membership inference attack resistance. To validate the practical value of the generated mobility, we assess its utility in enhancing downstream mobility prediction tasks. 

{\textbf{Uniqueness Testing.}} We randomly sample generated mobility sequences and compare their similarity to real mobility from the training dataset using Levenshtein distance~\cite{Levenshtein1966} as a metric~\cite{Cao2024STAGE,Yuan2024GenDailyActivities}. As shown in Figure~\ref{fig:privacy}, the majority of generated samples exhibit similarity scores between 0.4 and 0.8. This range indicates a balance: the generated data retains sufficient alignment with real-world patterns to ensure utility while maintaining enough diversity to avoid replicating any specific real individual’s mobility. The absence of a significant cluster of high similarity scores (above 0.8) further confirms that UniMob does not merely memorize training data but rather learns generalizable mobility patterns.

{\textbf{Memebership Inference Attacks.}}
This attack is designed to determine whether a given sample belongs to the training dataset~\cite{Cao2024STAGE,Yuan2024GenDailyActivities}. To reduce the influence of specific classification methods, we adopt three widely used classifiers: Random Forest (RF)~\cite{Breiman2001RF}, Multi-Layer Perceptron (MLP)~\cite{Rosenblatt1958MLP}, and Logistic Regression (LR)~\cite{Cox1958Logistic} to perform the attack. Results in Figure~\ref{fig:privacy} imply that attack success rates are consistently low across all classifiers, with no single model achieving a success rate above 0.65. This indicates that the generated mobility closely reflects the distribution of real data without retaining distinctive signatures of individual training samples, effectively resisting such privacy leakage. The consistent performance across classifiers strengthens the conclusion that UniMob’s design inherently reduces privacy risks, rather than relying on specific defense mechanisms against particular attack methods.

\begin{figure}[h]
    \centering
    \vspace{-2mm}
    \subfloat{\includegraphics[width=0.48\linewidth]{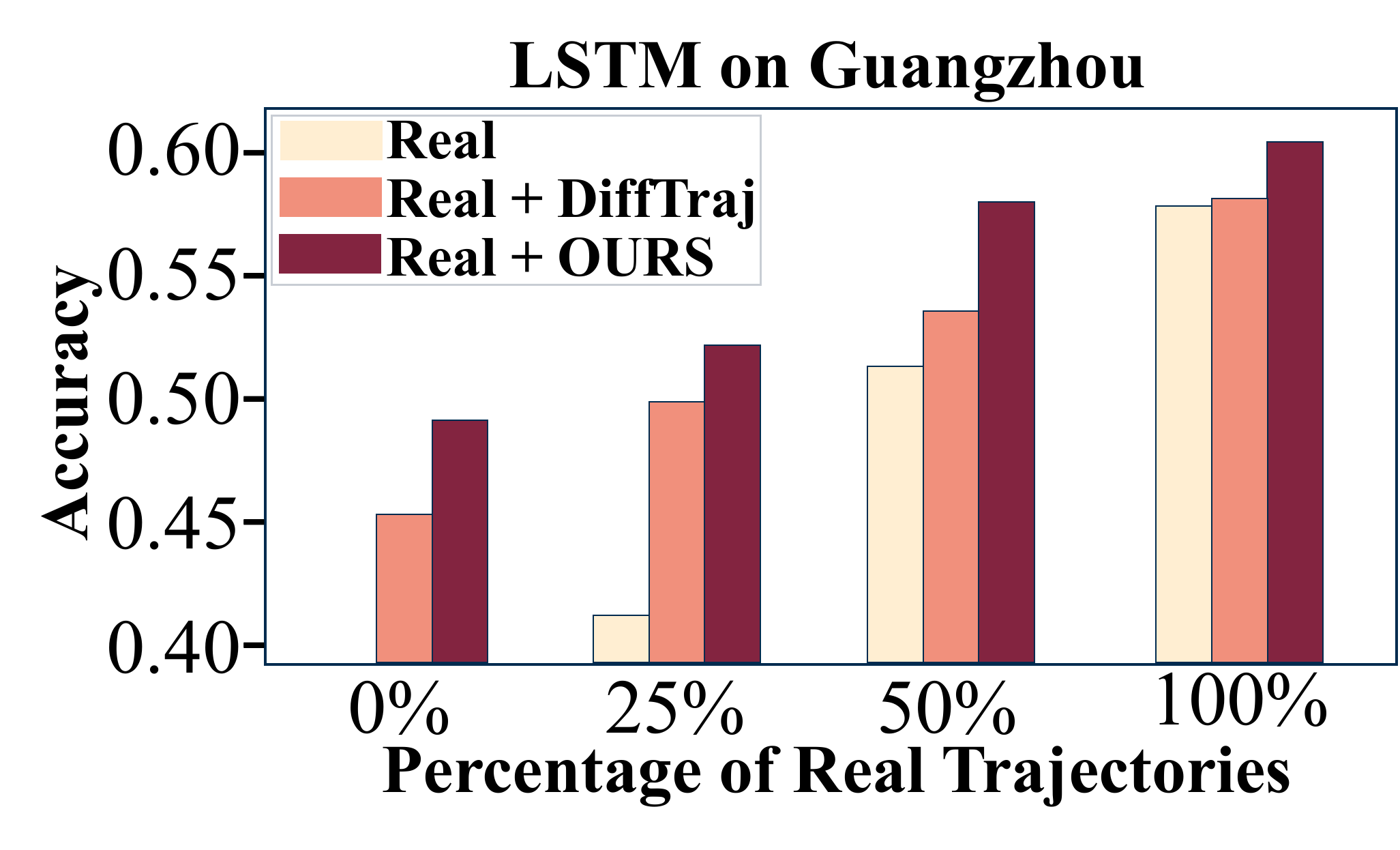}}
    \subfloat{\includegraphics[width=0.48\linewidth]{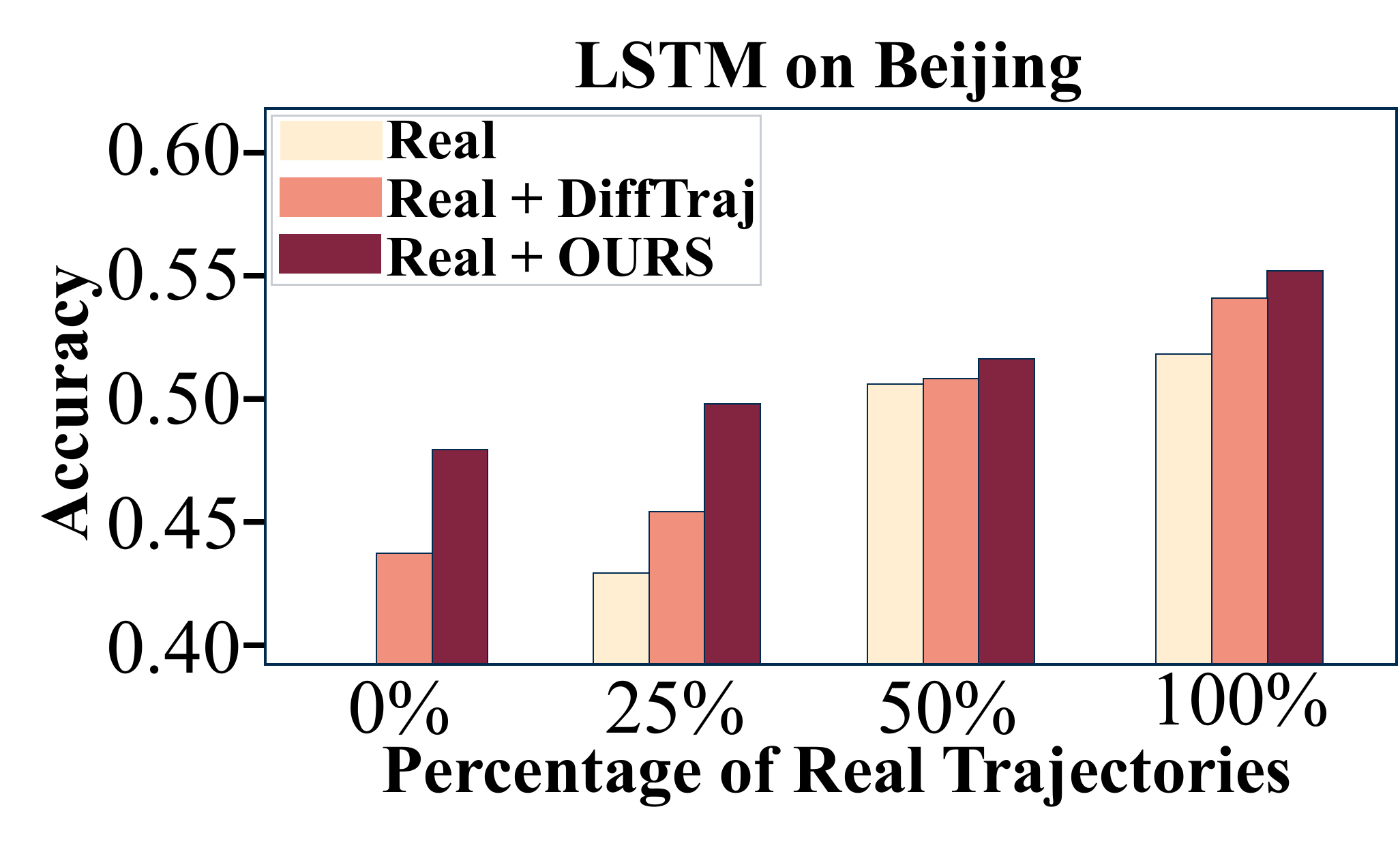}}\\
\vspace{-2mm}
    \subfloat{\includegraphics[width=0.48\linewidth]{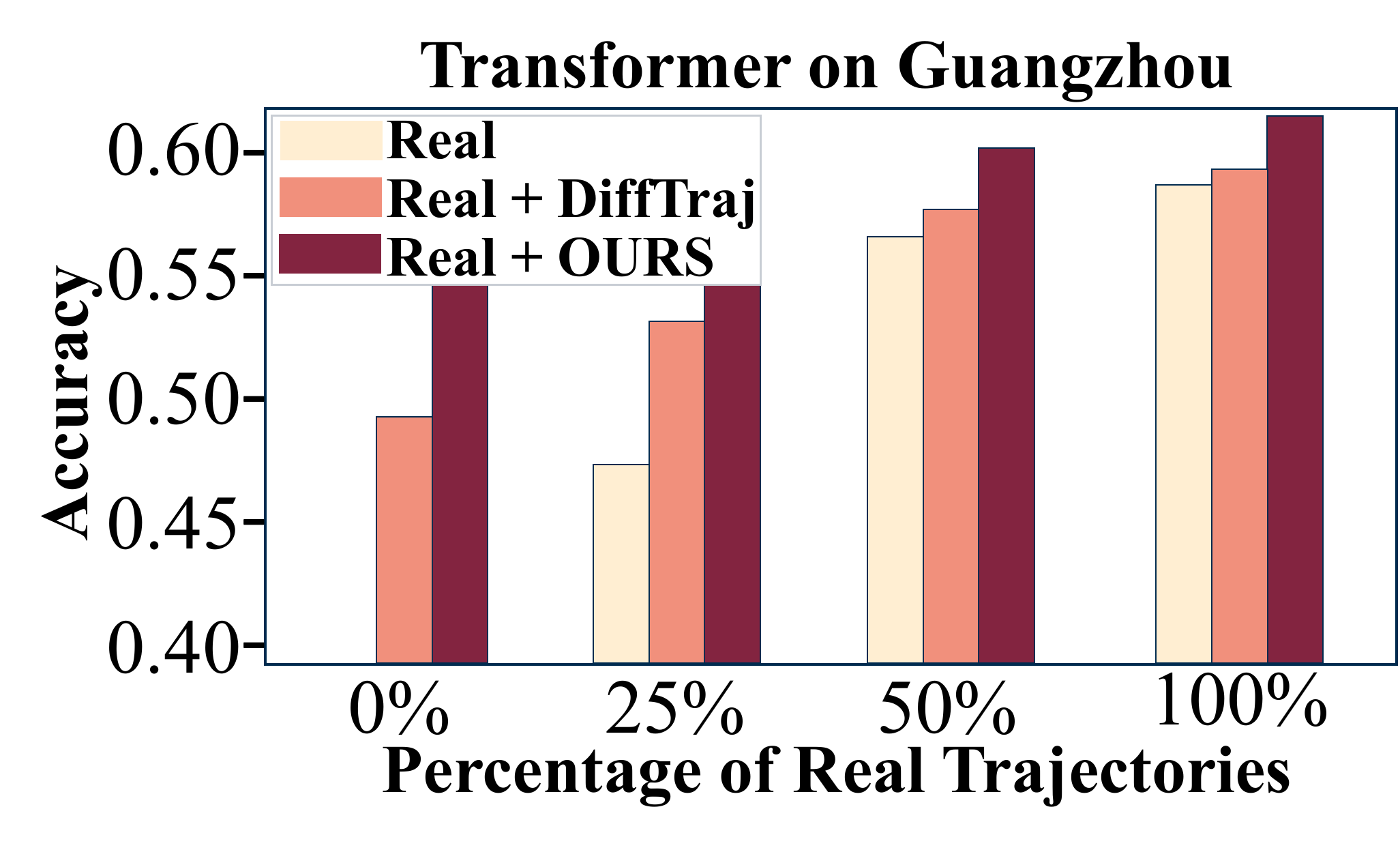}}
    \subfloat{\includegraphics[width=0.48\linewidth]{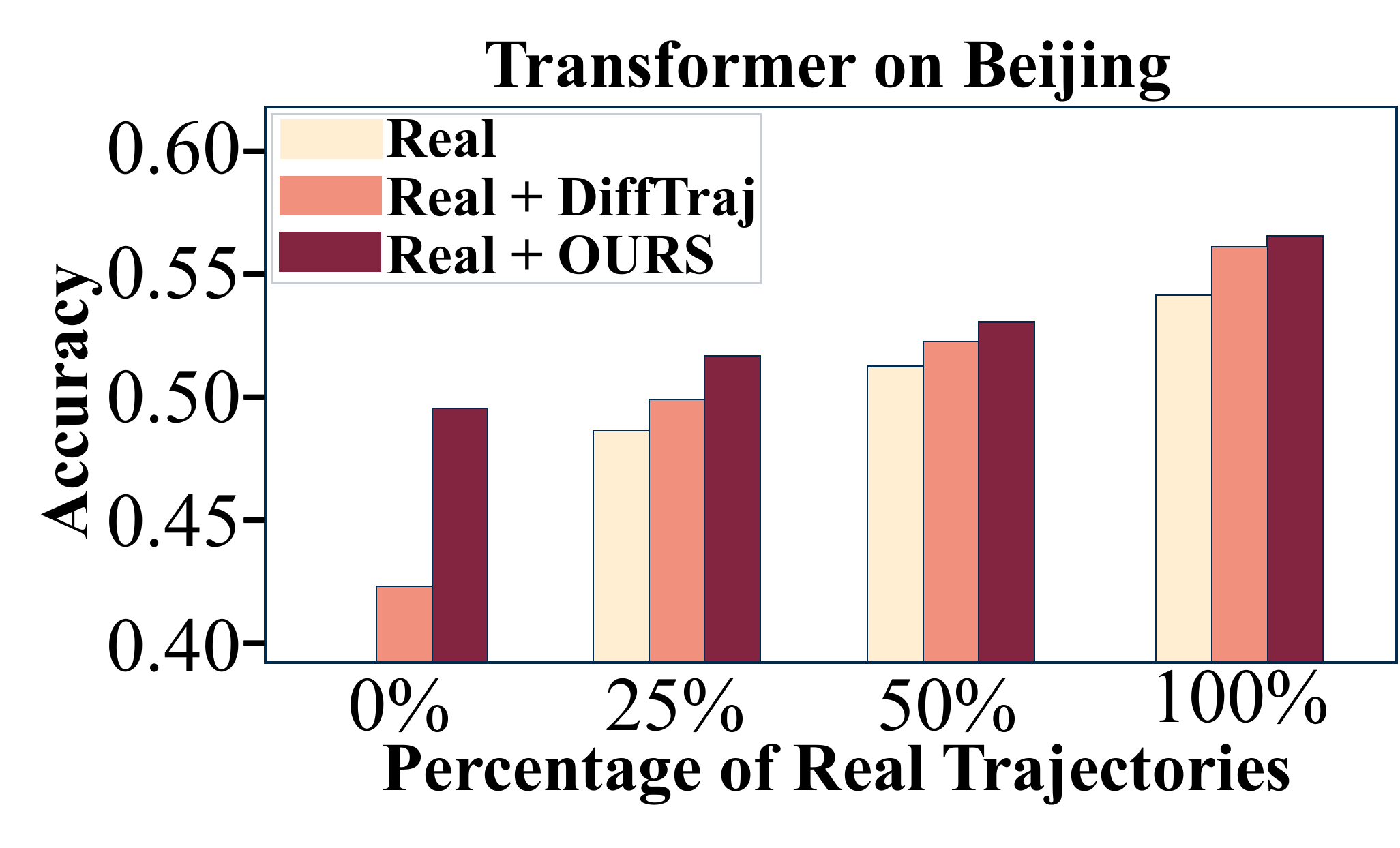}}
    \vspace{-2mm}
    \caption{Mobility prediction based on synthetic data.}
    \label{fig:utility of generated trajectories}
    \vspace{-2mm}
\end{figure}

{\textbf{Utility of the Generated mobility.}}
To evaluate the practical utility of our generated mobility, we assess its utility in enhancing downstream mobility prediction tasks.  We train LSTM- and Transformer-based models on mixed datasets containing varying proportions of real and synthetic mobility data, then evaluate their performance on real test samples. As shown in Figure~\ref{fig:utility of generated trajectories}, models trained with a combination of real data and UniMob-generated synthetic data consistently outperform those trained solely on real data or data augmented with synthetic samples from baselines like DiffTraj. This improvement is particularly pronounced when real data is scarce (e.g., 25\% real data), where the addition of UniMob-generated data boosts prediction accuracy by up to 8\% for LSTMs and 6\% for Transformers in both Beijing and Guangzhou datasets. Results confirm that UniMob-generated data is not merely privacy-preserving but also a valuable resource for addressing data scarcity in real-world applications.

These analyses demonstrate that UniMob generates mobility data that is both privacy-preserving, avoiding the risks of privacy leakage, and highly useful, enhancing the performance of downstream tasks even when real data is limited. This balance is a key advantage for deploying synthetic mobility in data-constrained urban environments.

\section{Related Work}
\label{related}
Human mobility generation methods has been extensively explored through two primary paradigms mechanism-driven and deep learning data-driven (DL) approaches. Mechanism-driven models~\cite{Jiang2016TimeGeo, Barbosa2015Recency} often rely on interpretable rules to simulate human mobility. A prominent example is the Exploration and Preferential Return (EPR) framework~\cite{Song2010Scaling}, where individuals probabilistically choose between revisiting known locations (preferential return) and exploring new ones sampled from distance‑decay distributions~\cite{pappalardo2016human}. Such models are valued for their simplicity and interpretability but are limited by rigid assumptions and lack flexibility in modeling complex spatiotemporal heterogeneity across cities. Recent advances in human mobility modeling have explored various generative methods to synthesize human mobility data. Early work, such as MoveSim~\cite{feng2020learning}, learned movement regularities from real-world mobility to simulate plausible user behaviors. TrajGen~\cite{zhang2022trajgen} adopted seq2seq and adversarial training to generate GPS trajectories with preserved utility and privacy. Choi et al.~\cite{choi2021trajgail} introduced TrajGAIL, leveraging generative adversarial imitation learning to replicate vehicle movement patterns in cities. Yuan et al.~\cite{yuan2022activity} proposed ActSTD, a neural ODE-based framework to model daily activity mobility with realistic timing and semantics. Long et al.~\cite{long2023practical} developed a two-stage variational model capturing user heterogeneity and trip-level variation. Jiang et al.~\cite{jiang2023continuous} designed TS-TrajGen, a GAN-based system incorporating A*-based priors to improve realism under road constraints. Existing data-driven methods merely perform well primarily in data-rich cities, while their effectiveness declines significantly in cities with limited data resources. Thus, a unified data-driven mobility generation model is urgently needed.
\section{Conclusion}
\label{conclusion}
In this paper, we propose \textbf{UniMob}, a unified generative framework for high-fidelity synthetic human mobility generation across cities. We introduce an LLM-based travel planner to infer high-level temporal and semantic travel plans, a unified spatial embedding to map varying spatial structures across cities into unified spatial representations, and a diffusion-based mobility generator to model the underlying spatiotemporal distribution of human mobility. Extensive experiments on two real-world human mobility datasets demonstrate that UniMob outperforms state-of-the-art baselines in human mobility generation across cities. Further analysis confirms its effectiveness in zero-shot and few-shot settings, the value of LLM guidance, its privacy-preserving capabilities, and its utility in downstream tasks. Our research sheds new light on the possibility of unified human mobility generation across cities, providing high-fidelity synthetic mobility data for diverse applications in both resource-rich and data-scarce urban environments.

\bibliographystyle{IEEEtran}
\bibliography{Reference}

\end{document}